\documentclass[apj]{emulateapj}

\usepackage{color}
\usepackage{amsmath}
\usepackage{booktabs}
\usepackage{enumerate}
\usepackage[version=3]{mhchem}
\bibpunct{(}{)}{;}{a}{}{,}

\shorttitle{}
\shortauthors{}

\begin{document}

\title{Chronology of Episodic Accretion in Protostars - an ALMA survey of the CO and H$_2$O snowlines}

\author{Tien-Hao Hsieh$^1$, Nadia M. Murillo$^2$, Arnaud Belloche$^3$, Naomi Hirano$^4$, Catherine Walsh$^5$, Ewine F. van Dishoeck$^{2,6}$, Jes K., J\o rgensen$^7$, Shih-Ping Lai$^{1,8}$}
\affiliation{$^1$Institute of Astronomy and Astrophysics, Academia Sinica, P.O. Box 23-141, Taipei 106, Taiwan}
\affiliation{$^2$Leiden Observatory, Leiden University, P.O. Box 9513, 2300 RA, Leiden, the Netherlands}
\affiliation{$^3$Max-Planck-Institut f\"{u}r Radioastronomie, Auf dem H\"{u}gel 69, 53121 Bonn, Germany}
\affiliation{$^4$Institute of Astronomy and Astrophysics, Academia Sinica, P.O. Box 23-141, Taipei 106, Taiwan}
\affiliation{$^5$School of Physics and Astronomy, University of Leeds, Leeds LS2 9JT, UK}
\affiliation{$^6$Max-Planck-Institut f\"{u}r extraterrestrische Physik, Giessenbachstra{\ss}e 1, 85748, Garching bei M\"{u}nchen, Germany}
\affiliation{$^7$Niels Bohr Institute, University of Copenhagen, {\O}ster Voldgade 5--7, DK 1350 Copenhagen K., Denmark}
\affiliation{$^8$Institute of Astronomy, National Tsing Hua University (NTHU), Hsinchu 30013, Taiwan}



\begin{abstract}
Episodic accretion has been used to explain the wide range of protostellar luminosities, but its origin and influence on the star forming process are not yet fully understood.
We present an ALMA survey of \ce{N2H+} ($1-0$) and \ce{HCO+} ($3-2$) toward 39 Class 0 and Class I sources in the Perseus molecular cloud.
 \ce{N2H+} and \ce{HCO+} are destroyed via gas-phase reactions with CO and \ce{H2O}, respectively, thus tracing the CO and \ce{H2O} snowline locations.
A snowline location at a much larger radius than that expected from the current luminosity suggests that an accretion burst has occurred in the past 
which has shifted the snowline outward.
We identified 18/18 Class 0 and 9/10 Class I post-burst sources from \ce{N2H+}, and 7/17 Class 0 and 1/8 Class I post-burst sources from \ce{HCO+}.
The accretion luminosities during the past bursts are found to be $\sim10-100~L_\odot$.
This result can be interpreted as either evolution of burst frequency or disk evolution.
In the former case, assuming that refreeze-out timescales are 1000 yr for \ce{H2O} and 10,000 yr for CO, 
we found that the intervals between bursts increases from 2400 yr in the Class 0 to 8000 yr in the Class I stage.
This decrease in the burst frequency may reflect that fragmentation is more likely to occur at an earlier evolutionary stage when the young stellar object is more prone to  instability.
\end{abstract}


\section{INTRODUCTION}
\label{sec:intro}
Episodic accretion plays an important role in star formation \citep{au14}.
In the episodic accretion scenario, a protostellar system stays in a quiescent accretion phase most of the time, and accretion bursts occasionally occur to deliver material onto the central protostar.
Because the accretion luminosity dominates the stellar luminosity at the early embedded phase \citep{ha96}, 
this behavior leads to a low protostellar luminosity for the majority of the time.
Such low luminosities have been revealed by recent surveys in star-forming regions with large statistical samples \citep{ev09,en09,kr12,hs13,du14}, yet 
it is still unknown how the accretion luminosity evolves and affects the luminosity distribution \citep{of11}.

The origin of episodic accretion is still unknown due to the difficulty in directly observing the accretion process.
The most plausible explanation is disk instability which can originate from several mechanisms such as thermal instability \citep{li85,be94}, gravitational instability \citep{vo05,vo10,bo08,ma11}, magnetorotational instability \citep{ar01,zh09,zh10a,zh10b}.
Stellar (or planet) encounters have also been proposed to explain accretion bursts \citep{cl96,lo04,fo10}.
Furthermore, \citet{pa14} suggest that the mass accretion rate is controlled by the mass infall from the large-scale turbulent cloud.
These possibilities make episodic accretion a key mechanism in star formation because it is associated with the timeline of disk fragmentation, planet formation, and multiplicity.
Observationally, \citet{li16} and \citet{ta18} found large-scale arms and arc structures in the disks toward four out of five FU Ori-type objects (\citealt{he66,he77}, see below) with $L_{\rm bol}\sim100-600~L_\odot$.
This supports the hypothesis that bursts are triggered by fragmentation due to gravitational instabilities. 
However, the remaining source, V1515 Cyg, hosts a smooth and symmetric disk.
From an evolutionary point of view, \citet{vo15} find that the outburst should preferentially occur during the Class I stage after the disk has accreted sufficient material to fragment.
However, \citet{hs18} found that accretion bursts with a few to a few tens of $L_\odot$ have occurred in Very Low Luminosity Objects (VeLLOs), which are extremely young or very-low mass protostars and thus unlikely hosts of massive disks.

Episodic accretion alters the star formation process by regulating the radiative feedback \citep{of09}.
The change in the thermal structure of the disk can directly affect the chemical composition of gas and ice \citep{ci16,wi19}. 
For example, \citet{ta16} found that complex organic molecules could be formed via gas-phase reactions in the hot region ($T\gtrsim$100 K) during the outburst phase.
In a continuous accretion process, the radiative feedback could suppress fragmentation by keeping the disk and/or cloud core warm \citep{of09,yi12,yi15,kr14}.
On the contrary, episodic accretion can moderate this effect, and during the quiescent phase, the disk has sufficient time to cool down and fragment \citep{st12}.
Such a process can be associated with the formation of binary/multiple systems and the formation of substellar objects, affecting the multiplicity and initial mass function \citep{kr10,st11,me16,ri18}.
Therefore, it is crucial to reveal the time intervals and the magnitude of outbursts in order to study how this radiative feedback affects the star formation process.

Variations in protostellar luminosity have been found in the past decades (i.e., \citealp[FU Orionis and EX Orionis events;][]{he66,he77}), which are considered to arise directly from episodic accretion.
Given the time intervals between bursts (\citealp[$\sim5\times10^{3}-5\times10^{4}$ yr,][]{sc13}), there are only a handful of cases in which luminosity variability has been reported to date (\citealp[V1647 Ori:][]{ab04,an04,ac07,fe07,as09}, \citealp[OO Serpentis:][]{ko07}, \citealp[\lbrack CTF93\rbrack 216-2:][]{ca11}, \citealp[VSX J205126.1:][]{co11,ko11}, \citealp[HOPS 383:][]{sa15}).
An outburst has also been detected toward the high-mass star forming region, S255IR-SMA1 \citep{ca16,li18}.
However, among these sources, HOPS 383 is the only source at an early stage (near the end of the Class I phase) due to 
the difficulty of infrared/optical observations to probe the embedded phase \citep{sa15}.
At longer wavelengths, \citet{lih18} found variations in millimeter flux of $30-60$\% toward 2 out of 29 sources using SMA.
The James Clerk Maxwell Telescope (JCMT) transient survey monitored 237 sources at submillimeter wavelengths for 18 months \citep{he17,jo18}, and they identified only one burst, from the Class I source, EC53.

\tabletypesize{\scriptsize}
\begin{deluxetable*}{ccccccccc}
\tabletypesize{\tiny}
\tablecaption{Targets}
\tablehead{ 	
\colhead{Source}	
& \colhead{Other name}	
& \colhead{R.A.}
& \colhead{Dec}	
& \colhead{$F_{\rm 1.2~mm}$}
& \colhead{$T_{\rm bol}$}	
& \colhead{$L_{\rm bol}$}	
& \colhead{P.A.}	
& \colhead{Reference}
\\
\colhead{}		
& \colhead{}
& \colhead{(J2000)}
& \colhead{(J2000)}	
& \colhead{mJy}
& \colhead{$T_\odot$}	
& \colhead{$L_\odot$}	
& \colhead{degree}	
& \colhead{}	
}
\startdata 
Per-emb-2	& IRAS 03292 + 3029	& 03h32m17.92s	& +30d49m47.85s	& 702.0$\pm$34.8	& 25	& 1.8	& 127	& 1,2,3\\
Per-emb-3	& 	& 03h29m00.58s	& +31d12m00.17s	& 52.6$\pm$1.1	& 30	& 0.9	& 277	& 1,4\\
Per-emb-4	& DCE065	& 03h28m39.11s	& +31d06m01.66s	& 0.9$\pm$0.2	& 28	& 0.3	& $^{*}$50	& 1,5\\
Per-emb-5	& IRAS 03282 + 3035	& 03h31m20.94s	& +30d45m30.24s	& 279.4$\pm$4.5	& 32	& 1.6	& 125	& 1,2,6\\
Per-emb-6	& DCE092	& 03h33m14.41s	& +31d07m10.69s	& 10.7$\pm$0.4	& 34	& 0.9	& 53	& 1,7\\
Per-emb-7	& DCE081	& 03h30m32.70s	& +30d26m26.47s	& 8.1$\pm$1.0	& 34	& 0.2	& 165	& 1,5\\
Per-emb-9	& IRAS 03267 + 3128, Perseus5	& 03h29m51.83s	& +31d39m05.85s	& 11.1$\pm$1.3	& 39	& 0.7	& 63	& 1\\
Per-emb-10	& 	& 03h33m16.43s	& +31d06m52.01s	& 21.9$\pm$0.6	& 26	& 1.4	& 230	& 1\\
Per-emb-14	& NGC 1333 IRAS4C	& 03h29m13.55s	& +31d13m58.10s	& 97.6$\pm$1.7	& 35	& 1.2	& 95	& 1,8\\
Per-emb-15	& RNO15-FIR	& 03h29m04.06s	& +31d14m46.21s	& 6.3$\pm$0.9	& 17	& 0.9	& 145	& 1,4\\
Per-emb-19	& DCE078	& 03h29m23.50s	& +31d33m29.12s	& 16.8$\pm$0.4	& 60	& 0.5	& 335	& 1,7\\
Per-emb-20	& L1455-IRS4	& 03h27m43.28s	& +30d12m28.78s	& 9.9$\pm$1.5	& 54	& 2.3	& 295	& 1\\
Per-emb-22	& L1448-IRS2	& 03h25m22.41s	& +30d45m13.21s	& 51.4$\pm$5.0	& 52	& 2.7	& 318	& 1,2,8\\
	& 	& 03h25m22.36s	& +30d45m13.12s	& -	& -	& -	& -\\
Per-emb-24	& 	& 03h28m45.30s	& +31d05m41.66s	& 3.9$\pm$0.3	& 62	& 0.6	& 281	& 1,8\\
Per-emb-25	& 	& 03h26m37.51s	& +30d15m27.79s	& 120.5$\pm$1.5	& 64	& 1.2	& 290	& 1,9,10\\
Per-emb-27	& NGC 1333 IRAS2A	& 03h28m55.57s	& +31d14m36.98s	& 247.6$\pm$12.4	& 54	& 30.2	& 204	& 1,2,4\\
	& 	& 03h28m55.57s	& +31d14m36.42s	& -	& -	& -	& -\\
Per-emb-29	& B1-c	& 03h33m17.88s	& +31d09m31.78s	& 133.1$\pm$5.5	& 41	& 4.8	& 110	& 1\\
Per-emb-30	& 	& 03h33m27.31s	& +31d07m10.13s	& 47.9$\pm$0.8	& 62	& 1.8	& 109	& 1,11\\
Per-emb-31	& DCE064	& 03h28m32.55s	& +31d11m05.04s	& 2.1$\pm$0.4	& 52	& 0.4	& 345	& 1,7\\
Per-emb-34	& 	& 03h30m15.17s	& +30d23m49.19s	& 11.4$\pm$0.5	& 93	& 1.9	& 45	& 1,10\\
Per-emb-35	& NGC 1333 IRAS1, Per-emb-35A	& 03h28m37.09s	& +31d13m30.76s	& 27.8$\pm$1.1	& 100	& 13.0	& 290	& 1,2,6\\
	& Per-emb-35B	& 03h28m37.22s	& +31d13m31.73s	& -	& -	& -	& -\\
Per-emb-36	& NGC 1333 IRAS2B	& 03h28m57.38s	& +31d14m15.74s	& 154.6$\pm$3.0	& 100	& 7.3	& 204	& 1,2,4\\
Per-emb-38	& DCE090	& 03h32m29.20s	& +31d02m40.75s	& 26.0$\pm$0.7	& 120	& 0.7	& 250	& 1,7\\
Per-emb-39	& 	& 03h33m13.82s	& +31d20m05.11s	& 2.0$\pm$0.4	& 59	& 0.1	& -	& 1\\
Per-emb-40	& B1-a	& 03h33m16.67s	& +31d07m54.87s	& 16.9$\pm$0.5	& 100	& 2.2	& 280	& 1,2\\
Per-emb-41	& B1-b	& 03h33m20.34s	& +31d07m21.32s	& 11.2$\pm$0.4	& 47	& 0.8	& 210	& 1,6\\
	& B1-bS	& 03h33m21.36s	& +31d07m26.37s	& -	& -	& -	& -\\
Per-emb-44	& SVS 13A, Per-emb-44-B	& 03h29m03.75s	& +31d16m03.77s	& 313.6$\pm$22.1	& 170	& 45.3	& 130	& 1,6\\
	& Per-emb-44-A	& 03h29m03.77s	& +31d16m03.78s	& -	& -	& -	& -\\
	& SVS 13A2	& 03h29m03.39s	& +31d16m01.58s	& -	& -	& -	& -\\
	& SVS 13B	& 03h29m03.08s	& +31d15m51.70s	& -	& -	& -	& -\\
Per-emb-45	& 	& 03h33m09.58s	& +31d05m30.94s	& 1.4$\pm$0.2	& 210	& 0.1	& -	& 1\\
Per-emb-46	& 	& 03h28m00.42s	& +30d08m00.97s	& 4.3$\pm$0.4	& 230	& 0.3	& 315	& 1\\
Per-emb-48	& L1455-FIR2	& 03h27m38.28s	& +30d13m58.52s	& 4.0$\pm$0.4	& 260	& 1.1	& 295	& 1\\
Per-emb-49	& Per-emb-49-A	& 03h29m12.96s	& +31d18m14.25s	& 21.9$\pm$1.9	& 240	& 1.4	& 207	& 1,6\\
	& Per-emb-49-B	& 03h29m12.98s	& +31d18m14.34s	& -	& -	& -	& -\\
Per-emb-51	& 	& 03h28m34.51s	& +31d07m05.25s	& 85.4$\pm$4.3	& 150	& 0.2	& 110	& 1\\
Per-emb-52	& 	& 03h28m39.70s	& +31d17m31.84s	& 4.3$\pm$0.4	& 250	& 0.2	& 25	& 1\\
Per-emb-54	& NGC 1333 IRAS6	& 03h29m01.55s	& +31d20m20.48s	& 3.1$\pm$0.4	& 230	& 11.3	& 310	& 1\\
Per-emb-58	& 	& 03h28m58.43s	& +31d22m17.42s	& 4.6$\pm$0.2	& 240	& 1.3	& 5	& 1\\
Per-emb-59	& 	& 03h28m35.06s	& +30d20m09.44s	& 1.6$\pm$0.1	& 49	& 0.5	& -	& 1\\
Per-emb-63	& 	& 03h28m43.27s	& +31d17m32.90s	& 24.6$\pm$0.5	& 490	& 2.2	& $^{*}$20	& 1,9\\
	& 	& 03h28m43.36s	& +31d17m32.69s	& -	& -	& -	& -\\
	& 	& 03h28m43.57s	& +31d17m36.31s	& -	& -	& -	& -\\
Per-emb-64	& 	& 03h33m12.85s	& +31d21m24.00s	& 39.7$\pm$0.6	& 480	& 4.0	& $^{*}$70	& 1\\
Per-emb-65	& 	& 03h28m56.32s	& +31d22m27.75s	& 35.6$\pm$0.7	& 440	& 0.2	& 140	& 1
\enddata
\tablecomments{
The source coordinates are obtained by a Gaussian fitting for the 1.2 mm images and the fluxes are listed in the next column (Figure \ref{fig:1.2mm}).
The bolometric temperature ($T_{\rm bol}$) and luminosity ($L_{\rm bol}$) are taken from \citet{du15}, in which $L_{\rm bol}$ is scaled to the new measured distance of Perseus (250 pc $\rightarrow$ 293 pc). We use $T_{\rm bol}=70~K$ \citep{ev09} as a boundary to classify Class 0 and Class I sources. 
The position angle (P.A.) of the outflow axis is taken from the corresponding references.\\
$^{*}$ The presumed P.A. of the source comes from disk or envelope structures rather than outflows. \\
References: (1) \citet{st18},(2) \citet{to16},(3) \citet{sc12},(4) \citet{pl13},(5) \citet{hs18},(6) \citet{le16},(7) \citet{hs17},(8) \citet{to15},(9) \citet{se18},(10) \citet{du14},(11) \citet{da08}
}
\label{tab:target}
\end{deluxetable*}

Chemical tracers are sensitive to the thermal history and can be used to probe past luminosity outbursts over a much longer timescale than direct observations of the luminosity \citep{le07,ki11,ki12,vi12,vi15}.
\citet{jo13} found a ring-like structure of H$^{13}$CO$^+$ surrounding the extended \ce{CH3OH} emission toward IRAS 15398-3359.
They found that this anti-correlation highlights the \ce{H2O} snowline location because \ce{CH3OH} has a sublimation temperature similar to that of \ce{H2O} and \ce{H2O} can destroy \ce{H^{13}CO+} via gas-phase reactions \citep{vi15}.
Given the current luminosity, they suggested that IRAS 15398-3359 has experienced a past luminosity outburst, sublimating \ce{H2O} over a larger region.
This result has later been confirmed with HDO by \citet{bj16} that directly revealed the radial extent of \ce{H2O} emission.
Such an anti-correlation between \ce{H2O} and \ce{H^{13}CO+} has also been found by \citet{va18b} in NGC 1333-IRAS2A, supporting that \ce{H^{13}CO+} is a good tracer of the water snowline.
An alternative method is to look directly at the CO snowline, which is at larger distances and can be more easily resolved using extended C$^{18}$O emission.
\citet{jo15} and \citet{fr17} studied 16 and 24 embedded protostars, respectively, and found that $20-50$ \% of the Class 0/I sources have experienced recent accretion bursts. 
Assuming a CO refreeze-out time of $\sim$10,000 yr, they estimated that the time interval between accretion bursts is $2-5\times10^{4}$ yr. 
Later, \citet{hs18} derived the CO snowline radius using the spatial anti-correlation of CO and N$_2$H$^+$ in VeLLOs, and found that 5 out of 7 sources are post-burst sources. On the other hand, using CO and N$_2$H$^+$ to trace the snowline, \citet{an16} found no evidence for past luminosity outbursts in four Class 0 protostars.

Here we present a survey of \ce{HCO+} and \ce{N2H+}, chemical tracers of past accretion bursts, in 39 protostars in Perseus that include 22 Class 0 sources and 17 Class I sources.  In Section \ref{sec:observations} we describe the sample and the observations.
The observational results are shown in Section \ref{sec:result}, and the detailed analysis and the modeling are given in Section \ref{sec:analysis}. Finally, we discuss the implications of our findings on episodic accretion, and summarize these results in Sections \ref{sec:discussion} and \ref{sec:summary}, respectively.

\section{Observations}
\label{sec:observations}
\subsection{Sample}
We selected 39 protostars from \citet{du15} in Perseus ($d=293$ pc, \citealt{or18}), a star-forming region containing sufficient Class 0 and Class I sources for a statistical survey.
These targets are located in the Western Perseus region \citep{hs13} mostly near the NGC1333 and B1 regions. 
Our sample includes 22 Class 0 and 17 Class I protostars with $T_{\rm bol}$ = $25-490$ K (Table \ref{tab:target}).
The bolometric luminosities were taken from \citet{du15} and were scaled with the new measured distance (250 pc $\rightarrow$ 293 pc), which yields $L_{\rm bol}$ = $0.1-45$ $L_\odot$.
These 39 sources are selected because they are detected at 850 or 1120 $\mu$m continuum emission by single dish observations (COMPLETE survey, \citealt{ri06});
the presence of the continuum emission suggests that the envelope has not yet dissipated, which might be a marker of stronger line emission.
All targets are included in the sample of the ``Mass Assembly of Stellar Systems and their Evolution with the SMA (MASSES)'' survey \citep{le15,le16,st18}, and we use the naming convention Per-emb-XX denoted by \citet{en09}.

\subsection{N$_2$H$^+$ ($1-0$) observation}
We observed the N$_2$H$^+$ ($1-0$) line emission toward 36 out of the 39 targets from 2018 March to 2018 April using ALMA (Cycle 5 project, 2017.1.01693.S, PI: T. Hsieh).
The N$_2$H$^+$ ($1-0$) data for the remaining three targets are taken from an earlier ALMA project (2015.1.01576.S), and the results of which were reported in \citet{hs18}.
With the C43-4 configuration, the resulting beam size was $\sim2\farcs4\times1\farcs5$ using natural weighting.
The largest scale covered is $\sim12\arcsec$.
The channel width was 30 kHz ($\sim$0.1 km s$^{-1}$ at a frequency of 93 GHz).
The on-source time toward each source was $\sim7.5$ min, resulting in an rms noise level of $\sim13$ mJy beam$^{-1}$ at a spectral resolution of 0.1 km s$^{-1}$.
The gain calibrator was J0336+3218 for all five executions.
The flux and bandpass calibrators were J0237+2848 for three executions and J0238+1636 for the remaining two executions.

\subsection{HCO$^+$ ($3-2$) and 1.2 mm continuum observations}
The HCO$^+$ ($3-2$) data were taken simultaneously with the continuum and CH$_3$OH ($2_{0,2}-1_{-1,1}$) data toward the 39 targets in 2018  September during the same project (2017.1.01693.S). 
The integration time is $\sim12$ min for each source.
The array configuration was C43-5, resulting in a beam size of $0\farcs45\times0\farcs30$ using natural weighting and $0\farcs33\times0\farcs22$ using uniform weighting; for each source, we choose the weighting based on the S/N of the image: uniform weighting is used for better detected sources. The largest scale covered is $\sim2\farcs6$.
The channel widths for both \ce{HCO+} and \ce{CH3OH} were 30.5 kHz ($\sim0.03~{\rm km~s^{-1}}$) and were averaged to 0.1 km s$^{-1}$ when imaging.
The rms noise level at a spectral resolution of 0.1 km s$^{-1}$ is $\sim8-13$ mJy beam$^{-1}$ depending on the weighting.
The window for continuum emission was centered at 268 GHz with a bandwidth of $\sim1.85$ GHz.
For all executions, the flux and band pass calibrators were J0237+2848, and the phase calibrator was J0336+3218.

\begin{figure*}
\centering 
\includegraphics[width=\textwidth]{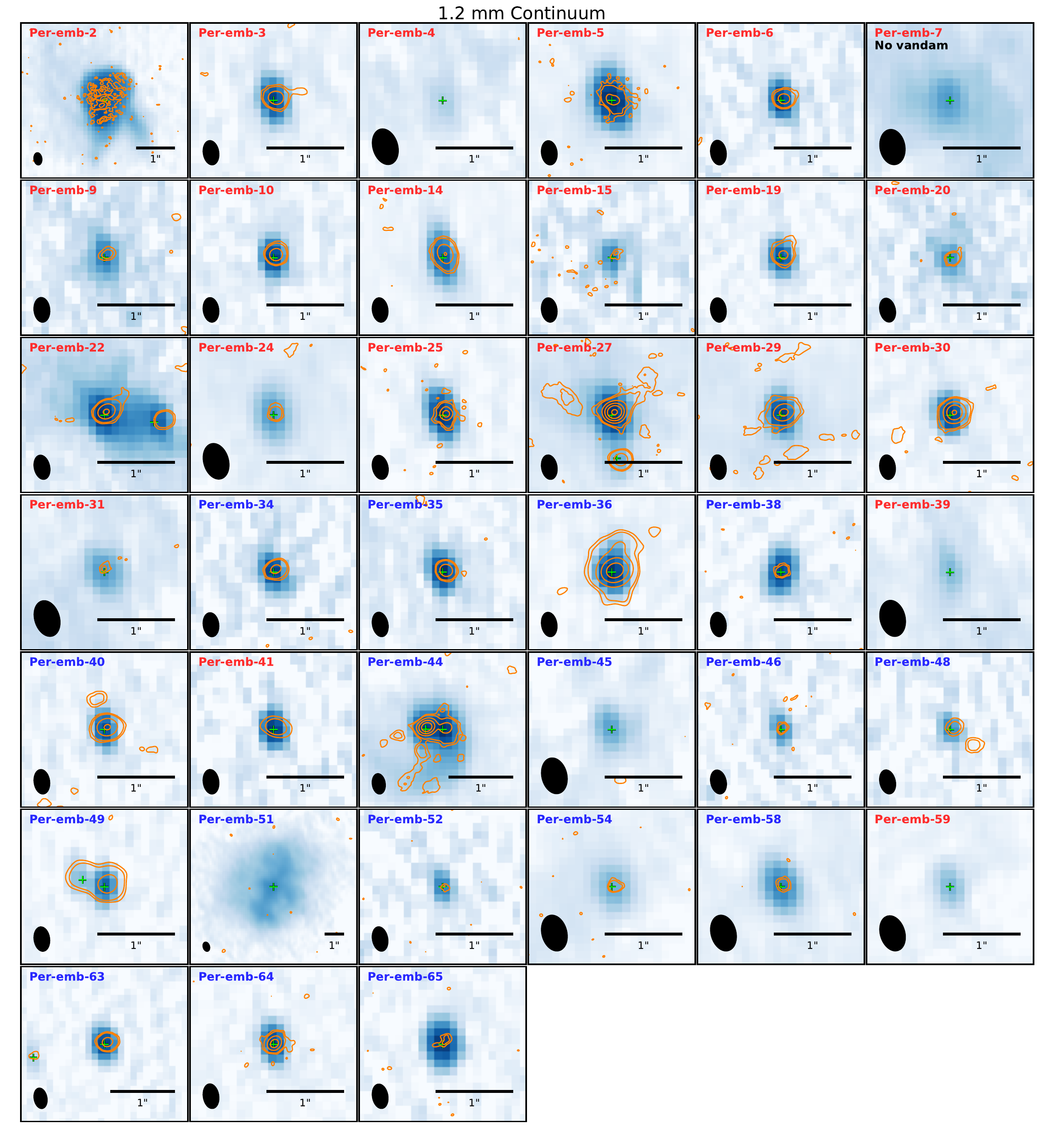}
\caption{1.2 mm dust continuum emission in color scale.
The color scales are artificially adjusted with the fluxes of each source listed in Table \ref{tab:target}.
The orange contours show the 8 mm continuum emission with the contour levels of 3$\sigma$, 5$\sigma$, 20$\sigma$, 50$\sigma$, 100$\sigma$, 200$\sigma$ from VANDAM \citep{to16}. 
The green cross indicates the position of the continuum emission from a Gaussian fitting.
The source names are labeled in the upper left corner, and the color indicates the stages of the sources as Class 0 (red) or Class I (blue).}
\label{fig:1.2mm}
\end{figure*}

\begin{figure*}
\centering 
\includegraphics[width=\textwidth]{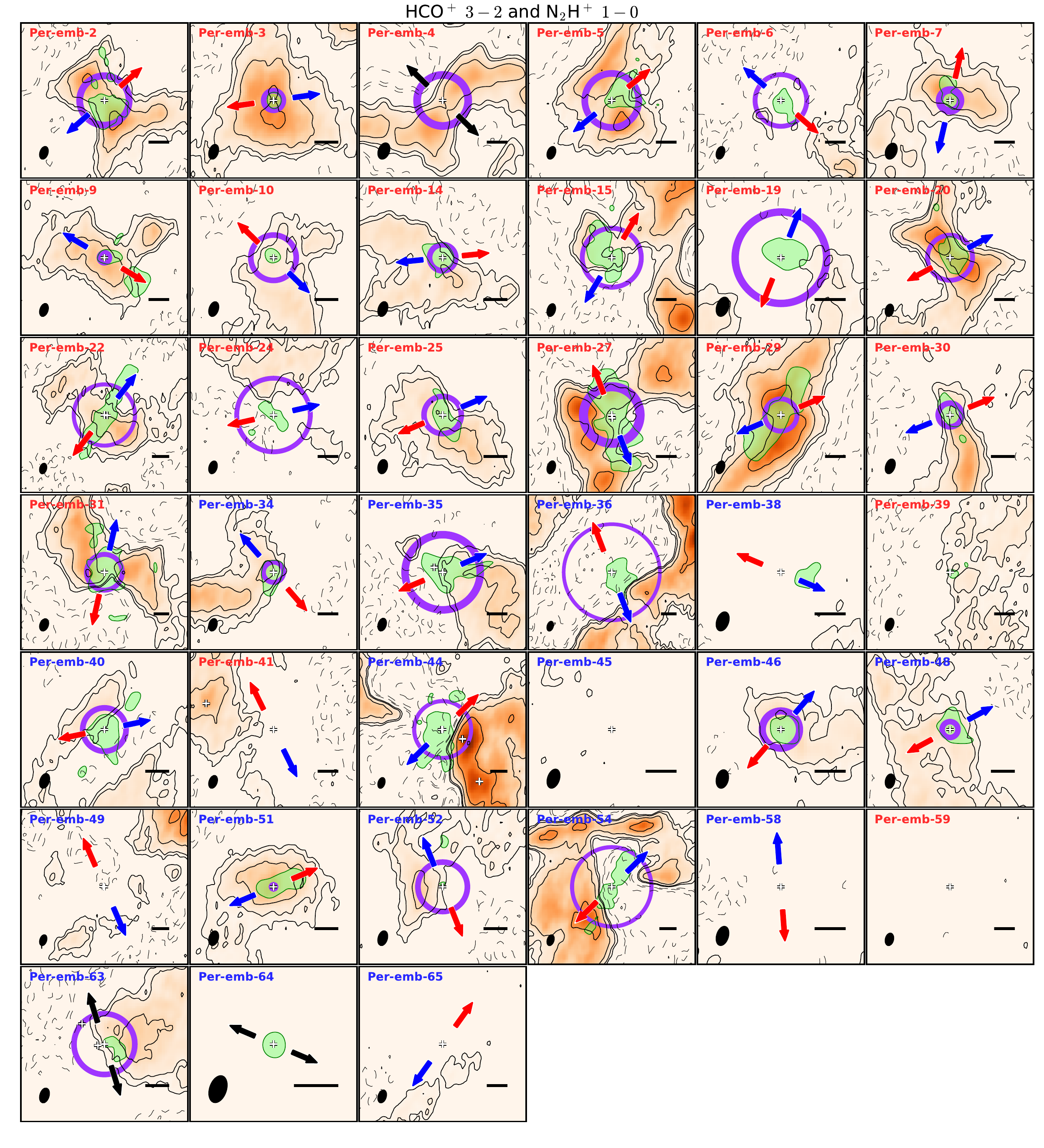}
\caption{Integrated intensity maps of N$_2$H$^+$ ($1-0$) emission in orange scale and contours.
The contour levels are 3$\sigma$, 5$\sigma$, 10$\sigma$, 20$\sigma$, and 40$\sigma$ with the rms noise level $\sigma$ listed in Table \ref{tab:map}.
The green area shows the \ce{HCO+} maps smoothed by a Gaussian kernel with $\sigma=0\farcs5$ at a noise level of 3$\sigma$ in order to compared with N$_2$H$^+$, and the original maps are shown in Figure \ref{fig:hcop}. The black bar in the lower right corner indicates a size of 4\arcsec while the size of the image is adjusted panel by panel. The red and blue arrows show the outflow directions from the literature, and the black arrows denotes the presumed outflow orientations from the disk or envelope structures. 
The white plus shows the positions of the continuum sources (Table \ref{tab:target}). The purple circles indicate the radii of the measured \ce{N2H+} peak, and their thicknesses represent the uncertainties (see the text for details).}
\label{fig:n2hp}
\end{figure*}

\section{Results}
\label{sec:result}
\subsection{Continuum emission at 1.2 mm}
Figure \ref{fig:1.2mm} shows the continuum emission at 1.2 mm of the sources in our sample.
These maps are centered at the source positions (as are the images in this paper) obtained from a Gaussian fitting to the continuum emission (Table \ref{tab:target}).
Table \ref{tab:target} lists also the Gaussian centers of the companion sources when detected. 
For these multiple systems, we use the coordinates from the brightest sources at 1.2 mm as the system centers.
The companions of Per-emb-40 and 48 found by the VLA Nascent Disk and Multiplicity Survey (VANDAM) at 8 mm \citep{to16} are not detected at 1.2 mm probably due to insufficient sensitivity.

\tabletypesize{\scriptsize}
\begin{deluxetable*}{ccccccccc}
\tabletypesize{\tiny}
\tablecaption{\ce{N2H+} and \ce{HCO+} integrated intensities}
\tablehead{ 
\colhead{}
& \multicolumn{3}{c}{\ce{N2H+}}
& \multicolumn{3}{c}{\ce{HCO+}}
\\ 
\cmidrule(lr){2-4} \cmidrule(lr){5-7}
\colhead{source}		
& \colhead{beam}		
& \colhead{vel. range}
& \colhead{rms}
& \colhead{beam}	
& \colhead{vel. range}
& \colhead{rms}	
\\
\colhead{}		
& \colhead{$''$}		
& \colhead{km s$^{-1}$}
& \colhead{mJy beam$^{-1}$ km s$^{-1}$}	
& \colhead{$''$}		
& \colhead{km s$^{-1}$}
& \colhead{mJy beam$^{-1}$ km s$^{-1}$}	
}
\startdata 
Per-emb-2	& $2.5\times1.5$	& $5.9-7.9$	& 25.4	& $0.33\times0.22$	& $3.9-6.1$, $7.9-9.3$	& 10.0\\
Per-emb-3	& $2.5\times1.5$	& $6.2-8.0$	& 24.1	& $0.33\times0.22$	& $5.1-6.4$, $7.8-9.5$	& 9.6\\
Per-emb-4	& $3.2\times2.0$	& $6.7-7.3$	& 9.8	& $0.46\times0.30$	& $5.9-8.0$	& 5.4\\
Per-emb-5	& $2.5\times1.5$	& $6.5-7.3$	& 21.3	& $0.34\times0.22$	& $5.0-6.4$, $8.2-9.5$	& 9.1\\
Per-emb-6	& $2.5\times1.5$	& $5.6-6.2$	& 16.3	& $0.43\times0.29$	& $5.1-5.4$, $7.8-8.2$	& 3.2\\
Per-emb-7	& $3.1\times2.0$	& $5.9-6.4$	& 14.3	& $0.33\times0.22$	& $3.1-5.5$, $6.6-8.4$	& 10.8\\
Per-emb-9	& $2.5\times1.5$	& $7.7-8.3$	& 17.8	& $0.34\times0.22$	& $5.0-7.6$, $8.6-10.2$	& 11.0\\
Per-emb-10	& $2.5\times1.5$	& $6.2-6.9$	& 19.3	& $0.34\times0.22$	& $4.6-5.6$, $7.4-8.7$	& 8.4\\
Per-emb-14	& $2.5\times1.5$	& $6.5-8.7$	& 25.3	& $0.34\times0.22$	& $5.0-6.7$, $8.7-10.3$	& 9.6\\
Per-emb-15	& $2.5\times1.5$	& $6.3-7.0$	& 19.3	& $0.38\times0.25$	& $2.6-5.9$, $8.7-9.0$	& 7.8\\
Per-emb-19	& $2.5\times1.5$	& $7.2-7.7$	& 18.5	& $0.38\times0.25$	& $6.0-6.9$, $8.0-8.8$	& 5.7\\
Per-emb-20	& $2.4\times1.5$	& $4.4-5.2$	& 21.0	& $0.33\times0.22$	& $2.5-4.1$, $6.2-8.3$	& 10.6\\
Per-emb-22	& $2.4\times1.5$	& $3.5-4.7$	& 23.0	& $0.33\times0.22$	& $1.2-3.3$, $4.9-7.9$	& 12.7\\
Per-emb-24	& $2.5\times1.5$	& $7.0-7.5$	& 17.6	& $0.38\times0.25$	& $5.6-6.6$, $8.1-9.1$	& 6.1\\
Per-emb-25	& $2.3\times1.5$	& $4.7-5.6$	& 20.8	& $0.33\times0.22$	& $2.7-4.6$, $6.1-8.1$	& 11.1\\
Per-emb-27	& $2.5\times1.5$	& $6.4-8.3$	& 25.5	& $0.34\times0.22$	& $4.7-5.9$, $8.8-10.8$	& 9.6\\
Per-emb-29	& $2.5\times1.5$	& $5.4-7.2$	& 24.4	& $0.34\times0.22$	& $3.9-5.7$, $7.3-8.9$	& 9.9\\
Per-emb-30	& $2.5\times1.5$	& $6.6-7.4$	& 19.9	& $0.34\times0.22$	& $-0.1-6.1$, $7.5-15.0$	& 22.2\\
Per-emb-31	& $3.2\times2.0$	& $6.7-7.6$	& 19.0	& $0.38\times0.25$	& $5.5-6.4$, $7.9-9.4$	& 7.0\\
Per-emb-34	& $2.4\times1.5$	& $5.6-6.4$	& 21.3	& $0.33\times0.22$	& $1.6-5.1$, $6.7-9.6$	& 14.5\\
Per-emb-35	& $2.5\times1.5$	& $6.8-7.4$	& 19.2	& $0.34\times0.22$	& $4.9-6.6$, $8.0-9.6$	& 10.5\\
Per-emb-36	& $2.5\times1.5$	& $7.0-7.6$	& 18.3	& $0.34\times0.22$	& $1.1-5.8$, $9.3-10.9$	& 13.6\\
Per-emb-38	& $2.5\times1.5$	& $6.3-6.7$	& 16.1	& $0.43\times0.29$	& $5.6-8.3$	& 6.4\\
Per-emb-39	& $2.5\times1.5$	& $6.6-7.3$	& 18.4	& $0.46\times0.30$	& $6.3-6.7$	& 2.4\\
Per-emb-40	& $2.5\times1.5$	& $5.7-7.2$	& 22.6	& $0.34\times0.22$	& $1.5-5.0$, $7.8-11.5$	& 14.8\\
Per-emb-41	& $2.5\times1.5$	& $6.4-6.9$	& 17.4	& $0.46\times0.30$	& $5.8-7.6$	& 5.2\\
Per-emb-44	& $2.5\times1.5$	& $7.7-9.1$	& 23.3	& $0.34\times0.22$	& $6.6-8.3$, $9.3-10.8$	& 9.9\\
Per-emb-45	& $2.5\times1.5$	& $6.7-7.3$	& 17.8	& $0.43\times0.29$	& $5.5-7.7$	& 5.7\\
Per-emb-46	& $2.4\times1.5$	& $4.7-5.3$	& 18.1	& $0.33\times0.22$	& $3.3-4.4$, $5.3-5.8$	& 7.0\\
Per-emb-48	& $2.4\times1.5$	& $3.9-5.2$	& 22.1	& $0.43\times0.28$	& $1.2-3.5$, $5.0-5.7$	& 6.8\\
Per-emb-49	& $2.5\times1.5$	& $8.1-8.7$	& 17.7	& $0.46\times0.30$	& $7.8-9.5$	& 4.9\\
Per-emb-51	& $2.4\times1.5$	& $6.5-7.2$	& 20.0	& $0.34\times0.22$	& $5.1-6.2$, $7.1-7.9$	& 7.7\\
Per-emb-52	& $2.5\times1.5$	& $7.6-8.2$	& 18.1	& $0.46\times0.30$	& $6.9-7.7$	& 3.4\\
Per-emb-54	& $2.5\times1.5$	& $7.7-8.4$	& 19.5	& $0.38\times0.25$	& $3.0-7.6$, $9.0-14.0$	& 14.0\\
Per-emb-58	& $2.5\times1.5$	& $7.6-8.4$	& 19.8	& $0.34\times0.22$	& $4.4-6.4$, $9.1-10.5$	& 10.4\\
Per-emb-59	& $2.4\times1.5$	& $5.2-5.8$	& 18.7	& $0.45\times0.30$	& $5.6-7.6$	& 5.7\\
Per-emb-63	& $2.5\times1.5$	& $7.5-8.3$	& 19.7	& $0.38\times0.25$	& $6.9-8.6$	& 5.6\\
Per-emb-64	& $2.5\times1.5$	& $4.1-4.6$	& 17.5	& $0.34\times0.22$	& $-3.1-6.1$, $7.7-15.0$	& 26.2\\
Per-emb-65	& $2.5\times1.5$	& $8.4-9.2$	& 21.0	& $0.46\times0.30$	& $6.5-9.1$	& 6.2
\enddata
\label{tab:map}
\end{deluxetable*}

\subsection{N$_2$H$^+$ maps}
Figure \ref{fig:n2hp} presents the N$_2$H$^+$ ($1-0$) integrated intensity maps and  
the velocity ranges over which they were integrated are listed in Table \ref{tab:map}.
In order to maximize the signal-to-noise ratio, we integrated the emission from all seven hyperfine components in N$_2$H$^+$ ($1-0$).
All targets are detected except for Per-emb-38, 45, 58, 59, and 64.
For Per-emb-41, the \ce{N2H+} emission is likely associated with B1b-S located in the north east.
The \ce{N2H+} emission in the Per-emb-39, 49, and 65 maps likely traces background cloud structures rather than the envelopes of the sources.
These non-detected and ambiguous sources are thus excluded in the following analysis.
As a result, 19 Class 0 and 11 Class I sources are remaining.
In most of the sources, the \ce{N2H+} emission peaks are offset from the continuum source, and are anti-correlated with \ce{HCO+} which is also shown in Figure \ref{fig:n2hp}.
This suggests that the \ce{N2H+} emission is suppressed in the warmer regions where it is destroyed through reactions with CO sublimating off dust grains \citep{ma91,jo04,va17}.
Several targets show strong negative contours in the map, likely caused by the spatial filtering of large-scale emission.
This is not expected to affect our analysis 
in which we measure the peak position of the \ce{N2H+} emission and compare it with model predictions.

\subsection{HCO$^+$ maps}
The blow-ups of the \ce{HCO+} integrated intensity maps are shown in Figure \ref{fig:hcop}.
We integrated the line emission excluding the optically thick region near the systemic velocity (Table \ref{tab:map}, see Appendix A). This effect of optical depth is discussed in section \ref{sec:opt}.
The velocity ranges of integration are listed in Table \ref{tab:map}.
We removed the following sources in the analysis because the emission does not seem to reflect the \ce{H2O} snowline in the envelope:
(1) Per-emb-4, 41, 49, and 59 show no detection near the source, (2) Per-emb-52 shows weak emission and ambiguous structures, and (3) For Per-emb-2, 36, 51, 64 and 65, the \ce{HCO+} emission peaks near the outflows axes, which are likely associated with the outflows rather than the envelopes.
We discuss if and how exclusion of these targets might introduce a bias.
Most of these targets (except for Per-emb-4, see section \ref{sec:peak} and \citealt{hs18}) in categories (1) and (2) have no \ce{N2H+} detections, and in category (3), Per-emb-36, 64, and 65 also show no (or ambiguous) \ce{N2H+} detections.
This suggests that their envelopes are almost dissipated; for Per-emb-36, 64, and 65 in category (3), outflows might dominate the emission with low envelope densities.
Since the dissipation is likely associated with the evolution rather than one accretion outburst, we speculate that the removal of these sources does not add a selection bias.
However, this narrows down our sampling evolutionary phase in the more evolved Class I stage with $T_{\rm bol}\gtrsim 250$ K.
Per-emb-2 (Class 0) and Per-emb-51 (Class I) from category (3) can introduce a bias but the small number only affects the statistical results by 5\% for the Class 0 stage and by 11\% for the Class I stage given the number of remaining sources.
We note that in order to minimize the outflow contamination, our analysis focusses on the emission roughly along the axis perpendicular to the outflow.
As a result, 18 Class 0 and 11 Class I sources are remaining for the following analysis.

\subsection{\ce{CH3OH} ($2_{0,2}-1_{-1,1}$) maps}
\begin{figure}
\centering 
\includegraphics[width=0.5\textwidth]{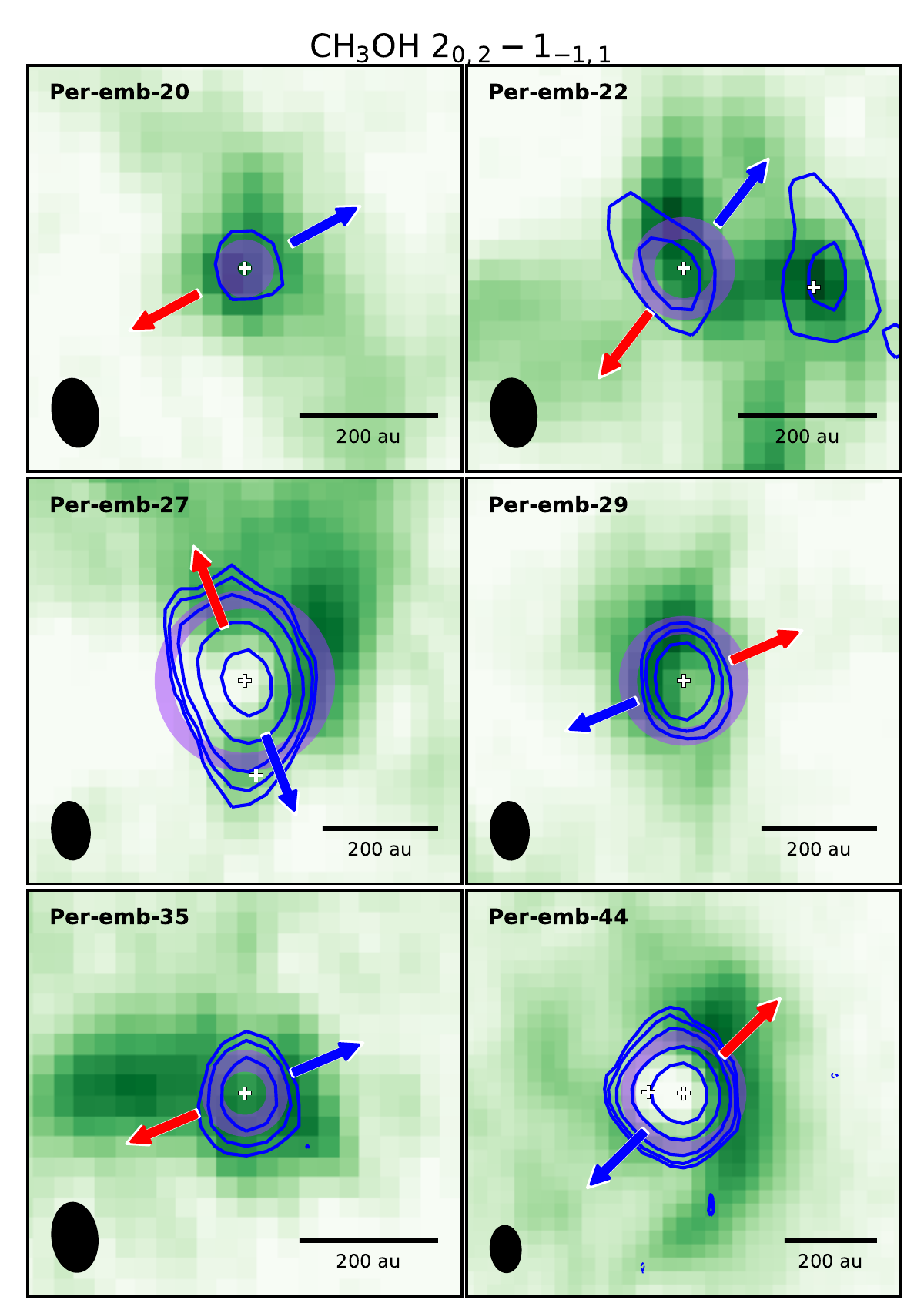}
\caption{\ce{CH3OH} integrated intensity map (blue contours) overlaid on that of \ce{HCO+} (green scale). The contour levels are 3$\sigma$, 5$\sigma$, 10$\sigma$, 30$\sigma$, and 70$\sigma$. The purple circles showing the radii of the measured \ce{HCO+} peak are the same as described in Figure \ref{fig:hcop}.}
\label{fig:ch3oh}
\end{figure}

CH$_3$OH ($2_{0,2}-1_{-1,1}$) at 254.015377 GHz is detected toward the source center in six targets (Figure \ref{fig:ch3oh}).
The emission most likely traces the region where \ce{CH3OH} sublimates due to central heating.
All these detected sources show an anti-correlation between \ce{CH3OH} and \ce{HCO+} except for Per-emb-20 and 22 which have very weak \ce{CH3OH} emission.
Because CH$_3$OH shares a similar sublimation temperature with H$_2$O ($\sim100$ K, \citealt{co04}), these anti-correlations could be used to confirm the radii of the H$_2$O snowlines \citep{va18}.
For Per-emb-20 and 22, \ce{CH3OH} and \ce{HCO+} emission have a similar peak position at the center (see also Appendix \ref{app:C}), which likely comes from unresolved structures at the current resolution.
However, the non-detections of \ce{CH3OH} do not indicate a temperature less than $\sim100$ K due to the unknown \ce{CH3OH} abundance and probably the low Einstein coefficient of $1.9\times10^{-5}~{\rm s^{-1}}$ for the transition.
The origin and the presence or absence of CH$_3$OH emission will be discussed in a separate paper (Murillo et al., in prep.).

\section{Analysis}
\label{sec:analysis}
In order to identify the post-burst sources, we model the line emission at different central luminosities.
We compare the \ce{N2H+} and \ce{HCO+} peak positions derived from the integrated intensity maps with those from the models.
Here we describe how we measured the peak positions from the observed images (Section \ref{sec:peak}) and how we construct the models (Section \ref{sec:mod}).
Then, we discuss the identification of sources that have likely experienced a past burst, i.e., post-burst sources, in section \ref{sec:postburst}, and the caveats in section \ref{sec:caveats}.

\subsection{The peak radii in the integrated intensity maps}
\label{sec:peak}
The difficulty in measuring the radius of the emission peak is that the observed peak position is not always located at the equatorial plane and such a plane is not necessarily perpendicular to the outflow. These misalignments can be reproduced by simulations \citep{of16} and are widely seen in observations \citep{le16,hs16,st18}.
To derive the radius of the peak emission, we employ a biconical mask to filter out the outflow contaminated region;
depending on sources, the region with a position angle smaller than $35-85\arcdeg$ from the outflow axis is excluded.
We identify the local maxima from the resulting maps as the peak position of the emission on one or two sides.
The mask in use and the selection of peak introduce artificial effects; taking the \ce{N2H+} emission in Per-emb-36 as an example, the measured radii of the emission peaks 
would be much smaller if the emission near the outflow axis were included in the analysis.
Thus, we carefully check the integrated maps in each source.
If only one peak is identified, the uncertainty is taken as the half-beam size.
If two peaks are found, the peak radius 
is defined as the average of their distances to the primary source and the uncertainty is the taken as the difference of that but with a minimum value as the half-beam size (Figures \ref{fig:n2hp}, \ref{fig:hcop}, and Table \ref{tab:sum}).

\begin{figure*}
\centering 
\includegraphics[width=0.8\textwidth]{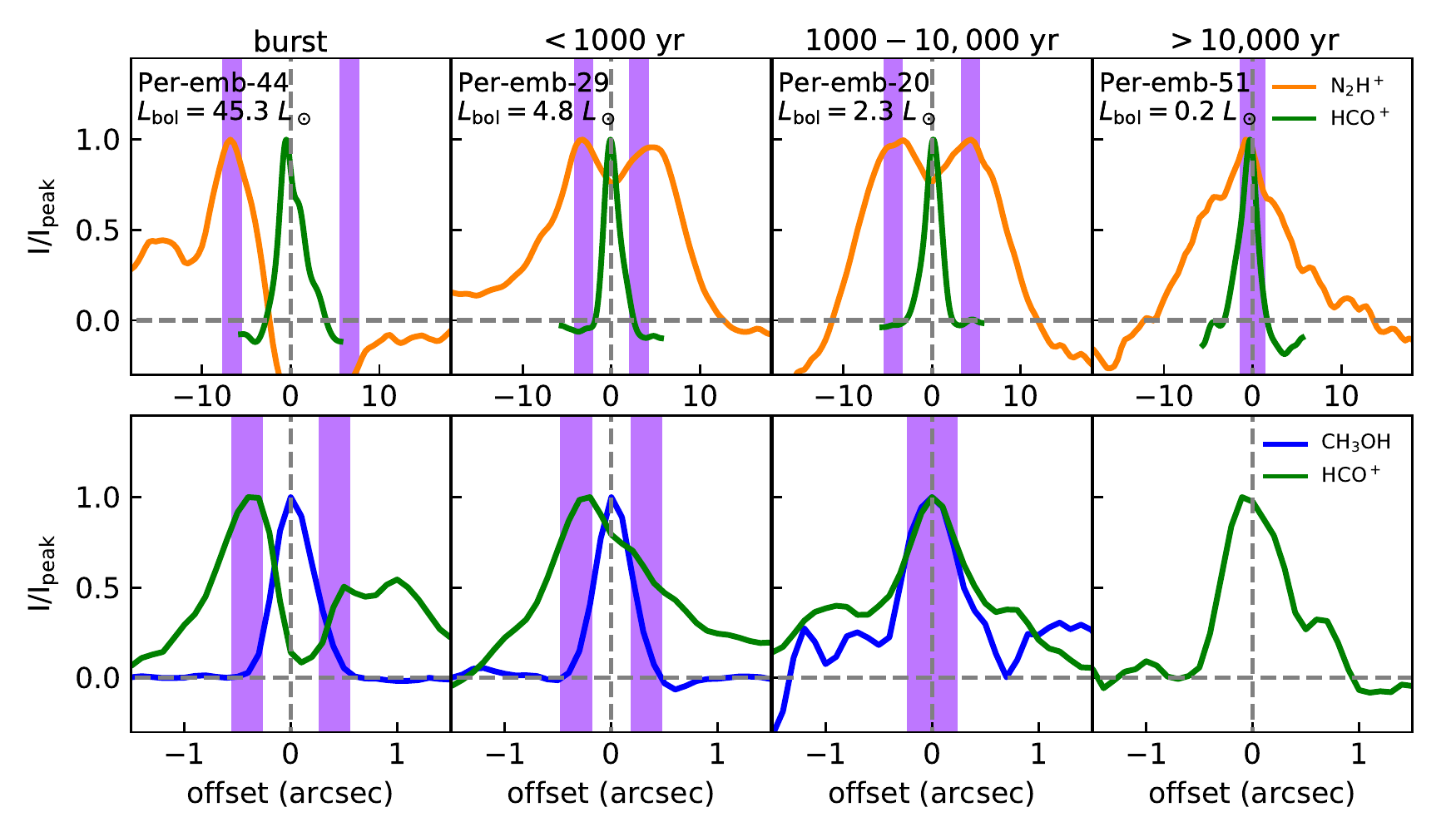}
\caption{ Intensity profiles along cuts across the source center and the local minima from the integrated intensity maps toward four sources.
The top panel shows the \ce{N2H+} and smoothed \ce{HCO+} profiles at large scales, and the bottom panel shows the \ce{HCO+} and \ce{CH3OH} profiles at small scales.
The profiles are normalized to their peaks.
The bolometric luminosity of the target is labeled in the upper left corner. The purple area indicates the radius of the measured peak for \ce{N2H+} (top) and \ce{HCO+} (bottom) in Table \ref{tab:sum}. The vertical and the horizontal dashed lines represent the source position and the intensity zero level, respectively.}
\label{fig:prof4}
\end{figure*}

We check using the images in Figure \ref{fig:n2hp} if the \ce{N2H+} peaks are located beyond the \ce{HCO+} emitting region; \ce{HCO+} is indeed expected to form from CO and traces the region where CO returns to the gas phase.
We find that most of the \ce{HCO+} emitting areas lie reasonably inside the \ce{N2H+} peak radii, at least along the major axis of the source.
For several sources, the elongated \ce{HCO+} emission is likely tracing the outflow and thus extends beyond the \ce{N2H+} peak radius (e.g. Per-emb-5, 7, 9, 22, 27, 29, 48, 51, and 54). 
This result suggests that the measured \ce{N2H+} peaks reflect the CO snowline radii. 
Per-emb-4 and 52 show \ce{N2H+} depletion without detections of \ce{HCO+} toward the center.
Per-emb-52 has extended \ce{C^{18}O} emission toward the center (Hsieh, T., in prep.).
On the other hand, the CO isotopologues, \ce{^{13}CO}, \ce{C^{18}O}, and \ce{C^{17}O} ($1-0$) and ($2-1$) are not (or marginally) detected in Per-emb-4 (\citealt{hs18}, DCE065 in the paper).
Thus, for Per-emb-4, we cannot exclude the possibility that \ce{N2H+} is absent due to freeze out of \ce{N2} in the central dense region \citep{be04}.

We check if the \ce{HCO+} peaks are located beyond the \ce{CH3OH} emitting regions for the six sources with \ce{CH3OH} detections (Figure \ref{fig:ch3oh}).
The spatial extent of \ce{CH3OH} emission is broadly within the radius of the measured \ce{HCO+} peak.
This suggests that \ce{HCO+} reflects well the \ce{H2O} sublimation region and that the measured radii are reasonable.
Unfortunately, most of the sources have no \ce{CH3OH} detection;
such non-detections do not rule out the hypothesis that \ce{HCO+} probes the location of the \ce{H2O} snowline but prevent us to confirm it.

Figure \ref{fig:prof4} shows the intensity profiles of \ce{N2H+}, \ce{HCO+}, and \ce{CH3OH} toward four standard sources along cuts across the source center and the identified local maxima. Anti-correlations are clearly seen in all plots except for \ce{N2H+}-\ce{HCO+} in Per-emb-51 and \ce{HCO+}-\ce{CH3OH} in Per-emb-20.
In the latter two cases, although these intensity profiles share a similar peak position, the \ce{HCO+} emission in Per-emb-51 and \ce{CH3OH} emission in Per-emb-20 are both very weak. The common peaks most likely come from an unresolved region smaller than the beam. The intensity profiles for all targets are shown in Appendix \ref{app:C}.

\begin{figure}
\centering 
\includegraphics[width=0.5\textwidth]{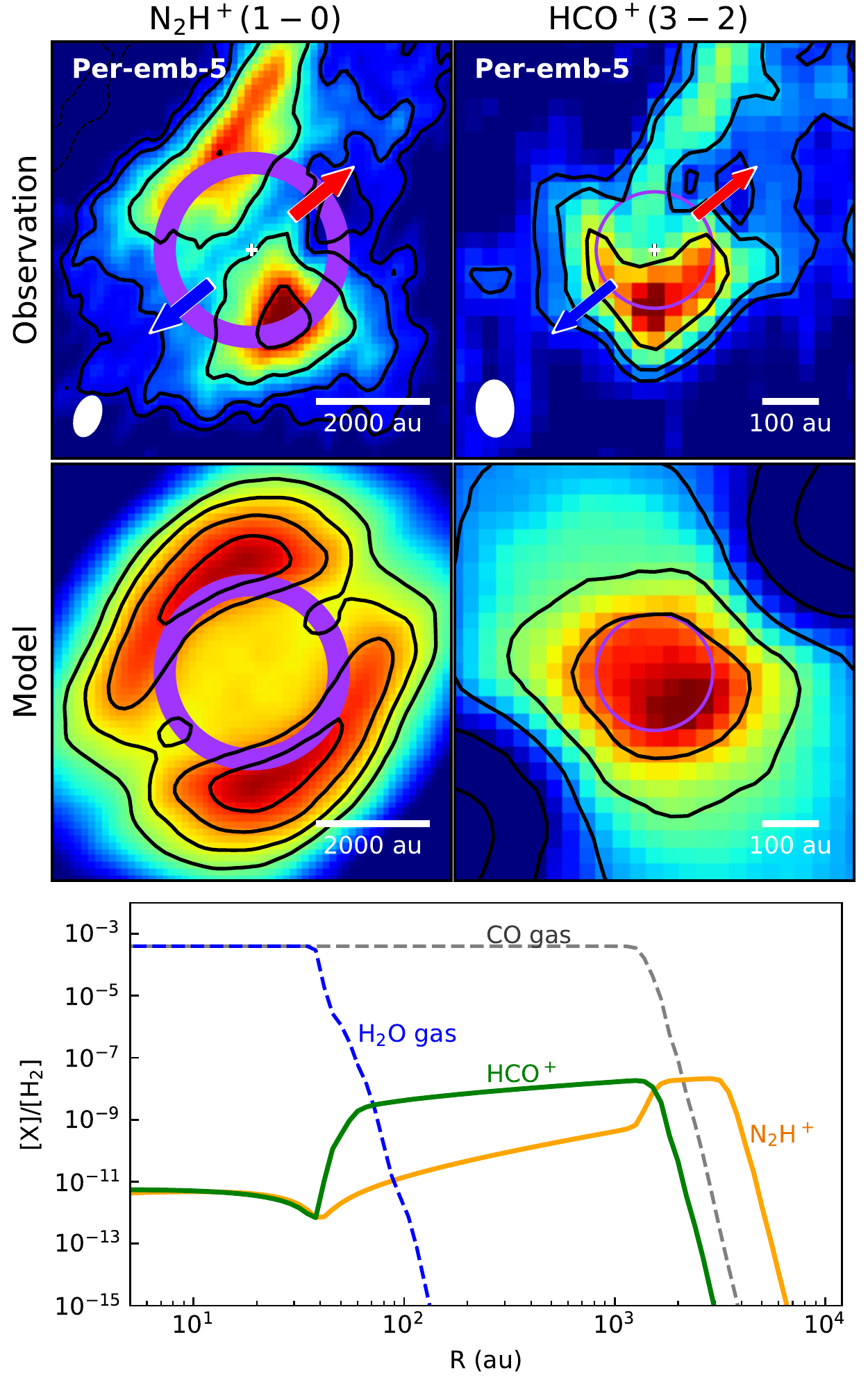}
\caption{An example showing the comparison of the observations and models for Per-emb-5. The top row shows the observed \ce{N2H+} map (left) and \ce{HCO+} map (right) in the same contours as in Figures \ref{fig:n2hp} and \ref{fig:hcop}. These images are at different sizes with scale bars shown in the bottom right corner. 
The images of a model with $30~L_\odot$ are shown in the second row with arbitrary intensity scales and contours to emphasize the emission peak. 
The purple circles indicate the radii of the \ce{N2H+} and \ce{HCO+} peaks in the corresponding map from observations.
The bottom plot shows the molecular abundance profiles in the equatorial plane for the model.}
\label{fig:mod}
\end{figure}

\begin{figure}
\centering 
\includegraphics[width=0.5\textwidth]{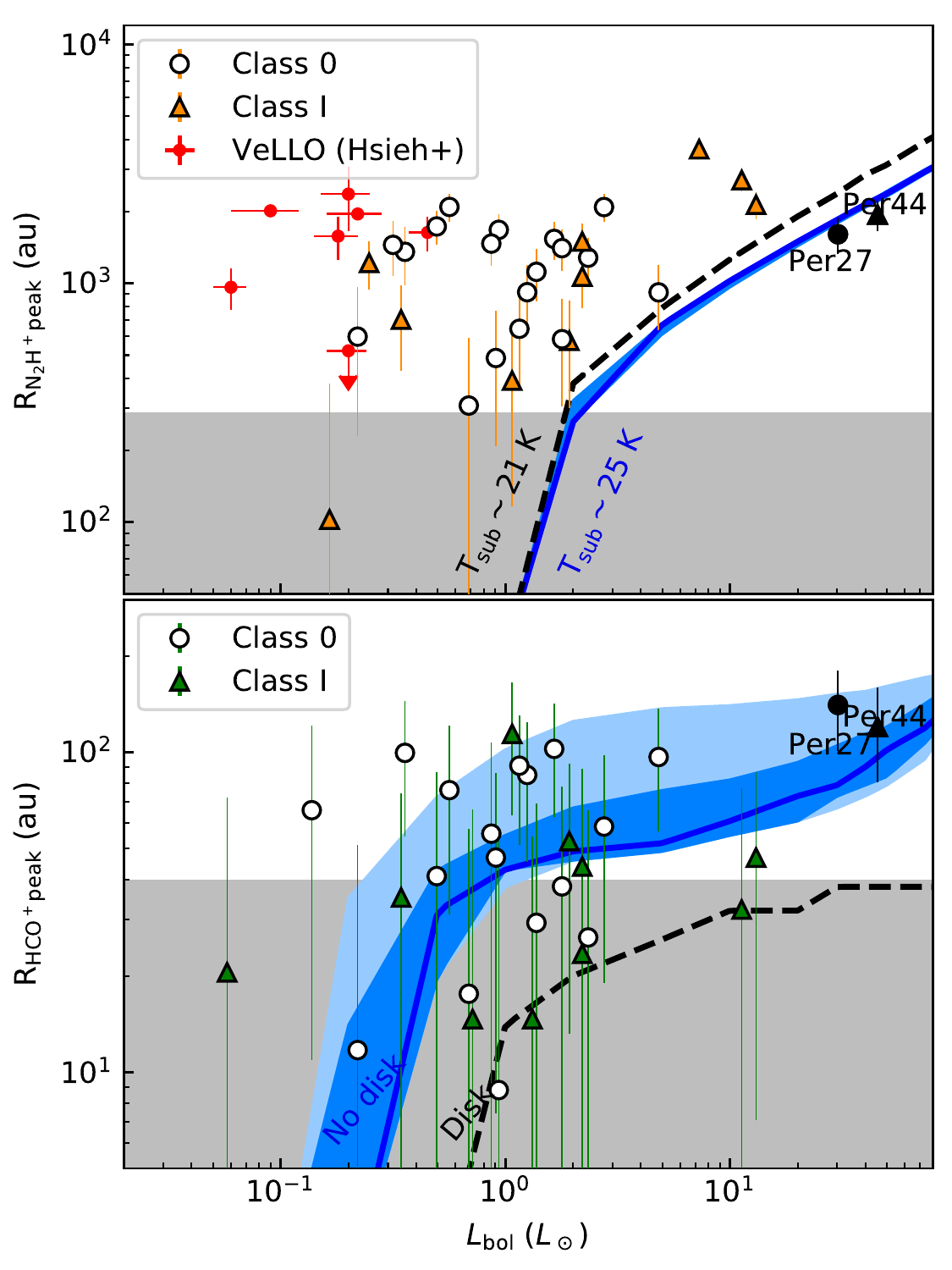}
\caption{Measured \ce{N2H+} (top) and \ce{HCO+} peak radii (bottom) as a function of the bolometric luminosities of the sources. 
The peak positions measured in this work are shown by open circles (Class 0) and filled triangles (Class I), and that from \citet{hs18} are in red.
Per-emb-27 and 44 have been classified as sources undergoing an accretion burst and are plotted in black.
We note that the peak position in \citet{hs18} is measured using abundance profiles, which could result in a slightly higher value. The modeled peak radius as a function of luminosity is shown with the blue lines for an inclination angle $\theta_{\rm inc}=45\arcdeg$. The dark and light blue area represent a range of $\theta_{\rm inc}=25-65\arcdeg$ and $\theta_{\rm inc}=15-75\arcdeg$, respectively, for which the upper boundary has a smaller inclination angle ($0\arcdeg$ for pole-on). The dashed black line shows a model at an inclination angle of $45\arcdeg$ with a CO sublimation temperature of 21 K (top) and with the presence of a disk (bottom). The grey area corresponds to a half-beam size, the limitation to resolve the depletion in the observations.}
\label{fig:lvsr}
\end{figure}

\subsection{MHW19 models}
\label{sec:mod}
To help the interpretation of the observational data, we construct a model framework for studying the relation between the luminosity and the emission peak radii of \ce{N2H+} and \ce{HCO+} (Murillo, N. M, in prep., hereafter MHW19).
MHW19 have built a grid of density structures by varying the disk and envelope geometry. 
For each density structure, RADMC3D\footnote{http://www.ita.uni-heidelberg.de/$\sim$dullemond/software/radmc-3d/} \citep{du12} is employed to calculate the temperature profile with a given central luminosity from 0.01 to 200 $L_\odot$. 
These physical conditions are later used for deriving the molecular abundance profiles.
The model starts from initially icy \ce{H2O} and \ce{CO}, and a chemical network based on the UMIST database for Astrochemistry, version RATE12 \citep{mc13} is used to calculate the static chemical abundance profiles.
The abundances of CO, \ce{N2} and \ce{H2O} relative to \ce{H2} are set to $2\times10^{-4}$, $1\times10^{-4}$, and $2\times10^{-4}$, respectively. 
A constant cosmic-ray ionization rate of appropriate for the interstellar medium, $1.2\times10^{-17}~{\rm s^{-1}}$, is adopted. 
It is noteworthy that \citet{pa16} found that the shocked gas from protostellar jets enhances the ionization rate by a few orders of magnitude, which can affect the abundances of \ce{HCO+} and \ce{N2H+} \citep{ga19}. However, this effect is less significant in the midplane of a flattened envelope which is the region in which we are interested.

Due to central heating, CO and \ce{H2O} are sublimated from the dust grains and destroy \ce{N2H+} and \ce{HCO+} via the gas-phase reactions
\begin{align*}
\ce{CO + N2H+ -> HCO+ + N2}
\end{align*}
and
\begin{align*}
\ce{H2O + HCO+ -> H3O+ + CO},
\end{align*}
respectively (see Section \ref{sec:intro}).
Thus, \ce{N2H+} and \ce{HCO+} are expected to be depleted in the central regions where the temperature is higher than the sublimation temperatures of CO ($\sim$20 K) and \ce{H2O} ($\sim$100 K), respectively.
Finally, MHW19 make line-emission images using the line radiative transfer code available as part of RADMC3D using molecular transition data from the Leiden database \citep{sh05} (\ce{N2H+} from \citealt{gr75} and \ce{HCO+} from \citealt{fl99}).
These images perform as a reference for comparison with observations with different burst luminosities $L_{\rm burst}$.

In this work, we adopt the model with a rotationally flattened envelope given by \citet{ul76}, i.e., without a disk component, from MHW19.
We include an outflow cavity with the edge, ${\rm z(r)}$ following the function ${\rm z(r)\propto r^{1.5}}$ with an opening angle of $50\arcdeg$ at ${\rm z=50~au}$ \citep{wh03,ro06}.
For a quantitative comparison between the model and observation, we measure the molecular peak radii from the models with different luminosities at different inclination angles (see Appendix \ref{app:B}).
Here we show an example comparing the observed \ce{N2H+} and \ce{HCO+} integrated intensity maps of Per-emb-5 to that of the model (Figure \ref{fig:mod});
the modeled image is generated using the CASA task ``simobserve'' for a source with a luminosity of 30 $L_\odot$ at an inclination angle of $45\arcdeg$.
Figure \ref{fig:lvsr} compares the modeled and observed molecular peak radii as a function of luminosity.
The modeled peak radius is affected by the inclination due to the emission contributed from the inner or outer envelope (see Appendix \ref{app:B}).
Besides, because a disk could shield the outer region from the central radiative heating, the modeled luminosity should be considered as a lower limit;
if a disk exists, a higher luminosity is needed to shift the peak position outward to match the observation (see the disk model in Figure \ref{fig:lvsr}).

\tabletypesize{\scriptsize}
\begin{deluxetable*}{ccccccccc}
\tabletypesize{\tiny}
\tablecaption{Identification of the past burst from \ce{N2H+} and \ce{HCO+}}
\tablehead{ 	
\colhead{name}	
& \colhead{$L_{\rm bol}$}	
& \colhead{$R_{\rm \ce{N2H+} peak}$}
& \colhead{$L_{\rm burst,\ce{CO}}$}	
& \colhead{$\dot{M}_{\rm acc}$}
& \colhead{$R_{\rm \ce{HCO+} peak}$}	
& \colhead{$L_{\rm burst,\ce{H2O}}$}
& \colhead{$\dot{M}_{\rm acc}$}
& \colhead{Last burst}
\\
\colhead{}		
& \colhead{$L_\odot$}
& \colhead{$''$}
& \colhead{$L_\odot$}	
& \colhead{$10^{-6}~M_\odot$ yr$^{-1}$}
& \colhead{$''$}	
& \colhead{$L_\odot$}
& \colhead{$10^{-6}~M_\odot$ yr$^{-1}$}
& \colhead{yr}
}
\startdata 
Per-emb-2	& 1.8	& $4.8\pm0.9$	& $18.1_{-0.1}^{+1.9}$	& $6.9_{-0.1}^{+0.7}$	& -	& -	& -	& $<10\,000$\\
Per-emb-3	& 0.9	& $1.7\pm0.9$	& $3.7_{-0.3}^{+0.1}$	& $1.4_{-0.1}^{+0.1}$	& $0.16\pm0.13$	& $1.6_{-1.0}^{+1.1}$	& $0.6_{-0.4}^{+0.4}$	& $<1000$\\
Per-emb-4	& 0.3	& $4.9\pm1.3$	& $20.0_{-1.9}^{+0.1}$	& $7.7_{-0.7}^{+0.1}$	& -	& -	& -	& $^*<10\,000$\\
Per-emb-5	& 1.6	& $5.2\pm1.0$	& $22.1_{-2.1}^{+0.1}$	& $8.5_{-0.8}^{+0.1}$	& $0.35\pm0.14$	& $49.2_{-22.2}^{+17.2}$	& $18.9_{-8.5}^{+6.6}$	& $<1000$\\
Per-emb-6	& 0.9	& $5.0\pm1.0$	& $20.0_{-0.1}^{+0.1}$	& $7.7_{-0.1}^{+0.1}$	& $0.19\pm0.18$	& $6.7_{-5.7}^{+4.3}$	& $2.6_{-2.2}^{+1.6}$	& $<1000$\\
Per-emb-7	& 0.2	& $2.0\pm1.3$	& $4.5_{-0.1}^{+0.1}$	& $1.7_{-0.1}^{+0.1}$	& $0.04\pm0.13$	& $<$0.9	& $<$0.3	& $1000-10\,000$\\
Per-emb-9	& 0.7	& $1.1\pm1.0$	& $2.2_{-0.2}^{+0.1}$	& $0.8_{-0.1}^{+0.1}$	& $0.06\pm0.14$	& $<$0.9	& $<$0.3	& $1000-10\,000$\\
Per-emb-10	& 1.4	& $3.8\pm1.0$	& $12.1_{-0.1}^{+1.3}$	& $4.6_{-0.1}^{+0.5}$	& $0.10\pm0.14$	& $<$0.9	& $<$0.3	& $1000-10\,000$\\
Per-emb-14	& 1.2	& $2.2\pm1.0$	& $4.9_{-0.1}^{+0.1}$	& $1.9_{-0.1}^{+0.1}$	& $0.31\pm0.14$	& $40.3_{-22.2}^{+14.1}$	& $15.4_{-8.5}^{+5.4}$	& $<1000$\\
Per-emb-15	& 0.9	& $5.7\pm0.9$	& $24.4_{-0.1}^{+0.1}$	& $9.4_{-0.1}^{+0.1}$	& $0.03\pm0.15$	& $<$1.6	& $<$0.6	& $1000-10\,000$\\
Per-emb-19	& 0.5	& $5.9\pm1.0$	& $27.0_{-0.1}^{+0.1}$	& $10.4_{-0.1}^{+0.1}$	& $0.14\pm0.16$	& $<$1.8	& $<$0.7	& $1000-10\,000$\\
Per-emb-20	& 2.3	& $4.4\pm0.9$	& $16.4_{-1.6}^{+0.1}$	& $6.3_{-0.6}^{+0.1}$	& $0.09\pm0.14$	& $<$0.9	& $<$0.3	& $1000-10\,000$\\
Per-emb-22	& 2.7	& $7.1\pm0.9$	& $36.4_{-0.1}^{+0.1}$	& $14.0_{-0.1}^{+0.1}$	& $0.20\pm0.14$	& $8.1_{-6.9}^{+6.7}$	& $3.1_{-2.6}^{+2.6}$	& $<1000$\\
Per-emb-24	& 0.6	& $7.1\pm0.9$	& $36.4_{-0.1}^{+0.1}$	& $14.0_{-0.1}^{+0.1}$	& $0.26\pm0.15$	& $24.4_{-19.5}^{+12.0}$	& $9.4_{-7.5}^{+4.6}$	& $<1000$\\
Per-emb-25	& 1.2	& $3.1\pm0.9$	& $9.0_{-0.9}^{+0.1}$	& $3.5_{-0.3}^{+0.1}$	& $0.29\pm0.13$	& $36.4_{-24.3}^{+12.8}$	& $14.0_{-9.3}^{+4.9}$	& $<1000$\\
Per-emb-27	& 30.2	& $5.5\pm1.0$	& $24.4_{-2.3}^{+0.1}$	& $9.4_{-0.9}^{+0.1}$	& $0.48\pm0.14$	& $99.1_{-25.7}^{+0.1}$	& $38.0_{-9.9}^{+0.1}$	& 0\\
Per-emb-29	& 4.8	& $3.1\pm1.0$	& $9.0_{-0.9}^{+0.1}$	& $3.5_{-0.3}^{+0.1}$	& $0.33\pm0.14$	& $44.5_{-22.4}^{+15.6}$	& $17.1_{-8.6}^{+6.0}$	& $<1000$\\
Per-emb-30	& 1.8	& $2.0\pm1.0$	& $4.5_{-0.4}^{+0.1}$	& $1.7_{-0.2}^{+0.1}$	& $0.13\pm0.14$	& $<$0.9	& $<$0.3	& $1000-10\,000$\\
Per-emb-31	& 0.4	& $4.6\pm1.3$	& $18.1_{-1.7}^{+0.1}$	& $6.9_{-0.7}^{+0.1}$	& $0.34\pm0.15$	& $49.2_{-24.8}^{+17.2}$	& $18.9_{-9.5}^{+6.6}$	& $<1000$\\
Per-emb-34	& 1.9	& $2.0\pm0.9$	& $4.0_{-0.1}^{+0.4}$	& $1.5_{-0.1}^{+0.2}$	& $0.18\pm0.14$	& $5.4_{-4.5}^{+2.7}$	& $2.1_{-1.7}^{+1.0}$	& $<1000$\\
Per-emb-35	& 13.0	& $7.3\pm0.9$	& $40.3_{-0.1}^{+0.1}$	& $15.4_{-0.1}^{+0.1}$	& $0.16\pm0.14$	& $1.6_{-1.0}^{+1.1}$	& $0.6_{-0.4}^{+0.4}$	& $1000-10\,000$\\
Per-emb-36	& 7.3	& $12.4\pm0.9$	& $109.5_{-0.1}^{+0.1}$	& $42.0_{-0.1}^{+0.1}$	& -	& -	& -	& $<10\,000$\\
Per-emb-38	& 0.7	& -	& -	& -	& $0.05\pm0.18$	& $<$6.7	& $<$2.6	& $>1000$\\
Per-emb-39	& 0.1	& -	& -	& -	& $0.23\pm0.19$	& $13.4_{-11.6}^{+11.0}$	& $5.1_{-4.4}^{+4.2}$	& $<1000$\\
Per-emb-40	& 2.2	& $3.6\pm1.0$	& $11.0_{-0.1}^{+1.2}$	& $4.2_{-0.1}^{+0.5}$	& $0.08\pm0.14$	& $<$0.9	& $<$0.3	& $1000-10\,000$\\
Per-emb-41	& 0.8	& -	& -	& -	& -	& -	& -	& -\\
Per-emb-44	& 45.3	& $6.6\pm1.0$	& $33.0_{-0.1}^{+0.1}$	& $12.7_{-0.1}^{+0.1}$	& $0.41\pm0.14$	& $73.4_{-24.2}^{+7.7}$	& $28.1_{-9.3}^{+3.0}$	& 0\\
Per-emb-45	& 0.1	& -	& -	& -	& $0.07\pm0.18$	& $<$6.7	& $<$2.6	& $>1000$\\
Per-emb-46	& 0.3	& $2.4\pm0.9$	& $5.4_{-0.1}^{+0.1}$	& $2.1_{-0.1}^{+0.1}$	& $0.12\pm0.13$	& $<$0.9	& $<$0.3	& $1000-10\,000$\\
Per-emb-48	& 1.1	& $1.3\pm0.9$	& $2.7_{-0.3}^{+0.3}$	& $1.0_{-0.1}^{+0.1}$	& $0.39\pm0.17$	& $66.4_{-26.1}^{+14.7}$	& $25.5_{-10.0}^{+5.6}$	& $<1000$\\
Per-emb-49	& 1.4	& -	& -	& -	& -	& -	& -	& -\\
Per-emb-51	& 0.2	& $0.3\pm1.0$	& $<$2.2	& $<$0.8	& -	& -	& -	& $>10\,000$\\
Per-emb-52	& 0.2	& $4.2\pm0.9$	& $14.8_{-1.4}^{+0.1}$	& $5.7_{-0.5}^{+0.1}$	& -	& -	& -	& $<10\,000$\\
Per-emb-54	& 11.3	& $9.2\pm0.9$	& $60.1_{-0.1}^{+6.3}$	& $23.0_{-0.1}^{+2.4}$	& $0.11\pm0.15$	& $<$1.6	& $<$0.6	& $1000-10\,000$\\
Per-emb-58	& 1.3	& -	& -	& -	& $0.05\pm0.14$	& $<$0.9	& $<$0.3	& $>1000$\\
Per-emb-59	& 0.5	& -	& -	& -	& -	& -	& -	& -\\
Per-emb-63	& 2.2	& $5.1\pm0.9$	& $20.0_{-0.1}^{+2.1}$	& $7.7_{-0.1}^{+0.8}$	& $0.15\pm0.15$	& $<$1.6	& $<$0.6	& $1000-10\,000$\\
Per-emb-64	& 4.0	& -	& -	& -	& -	& -	& -	& -\\
Per-emb-65	& 0.2	& -	& -	& -	& -	& -	& -	& -\\
\enddata
\tablecomments{
Col. (1): Source name.
Col. (2): Bolometric luminosity from \citet{du14} scaled from 230 pc to 293 pc.
Col. (3): Radius of the measured \ce{N2H+} peak.
Col. (4): Luminosity corresponding to the radius of the measured \ce{N2H+} peak from models, i.e., the accretion luminosity during the past burst.
Col. (5): Mass accretion rate estimated from Col. (4).
Col. $(6)-(8)$: Same as Col. $(3)-(5)$ but with the numbers measured from \ce{HCO+}.
Col. (9): Time after the last burst.
$^{*}$ Per-emb-4 has no \ce{HCO+} and CO detection toward the center. Thus, it is unclear if the \ce{N2H+} depletion comes from destruction by CO or freeze-out of the parent molecule, \ce{N2}. 
}
\label{tab:sum}
\end{deluxetable*}

\subsection{Identification of post-burst sources}
\label{sec:postburst}
Figure \ref{fig:lvsr} shows the measured radii of \ce{N2H+} and \ce{HCO+} emission peaks as a function of the source bolometric luminosity. 
If the observed peak radius is larger than the predicted value from the model at the given luminosity, 
the source has likely experienced a past accretion burst, i.e., it is a post-burst source \citep{le07,jo13}.
After the burst, the refreeze out of CO and \ce{H2O} should start from the inner high-density region such that the observed peak radii are likely static \citep{le07,vi15,hs18}.
Therefore, comparing these peak radii with the model, we can estimate the peak luminosity in the past, i.e., the burst luminosity ($L_{\rm burst}$, Table \ref{tab:sum}), and identify the post-burst sources with $L_{\rm burst}>L_{\rm bol}$.
However, this estimated luminosity is degenerate with the inclination angle for the case of \ce{HCO+} at small scales (Figure \ref{fig:lvsr}).
This degeneracy becomes severe near the pole-on case (see Appendix \ref{app:B}).
Fortunately, most of our targets show clear bipolar outflows \citep{st18}, suggesting that they are not pole-on sources. 
Statistically, the nearly pole-on probability ($\theta_{\rm inc}<25\arcdeg$) is less than 10\%.
Thus, we derive the burst luminosity by comparing the measured peak radius to the model at an inclination angle of 45$^\circ$ and use the angle from $25\arcdeg-75\arcdeg$ as the uncertainty (Table \ref{tab:sum}).

As a result, we found that with \ce{N2H+}, almost all Class 0 and Class I sources are identified as post-burst sources. With \ce{HCO+}, 10/17 Class 0 sources and 2/10 Class I sources are identified as post-burst sources.
The sources with a peak radius less than the half-beam size are not classified as post-burst sources;
they are classified as sources without a past burst or sources where CO or \ce{H2O} have refrozen onto the dust grains after the last burst.

It is noteworthy that Per-emb-4 has no detection of \ce{HCO+} nor CO isotopologues \citep{hs18} in the \ce{N2H+} depletion region.
Thus, we are not able to exclude the possibility that the \ce{N2H+} depletion comes from freeze out of its parent molecule, \ce{N2}.

\subsection{Caveat in identification of the post-burst sources}
\label{sec:caveats}
\subsubsection{Effect of optical depth for probing the \ce{H2O} snowline}
\label{sec:opt}
The optical depth from the continuum or line emission could affect the measured peak positions. 
If the line emission and dust continuum emission are optically thin, the integrated intensity maps should properly reflect the snowline locations.
\ce{N2H+} is expected to be less affected by this issue, because it is usually optically thin throughout the outer envelope due to its relatively low abundance.
However, for the \ce{H2O} snowline traced by \ce{HCO+} in the inner dense region, the effects of optical depth need to be addressed.
Here we discuss how it influences the measured snowline radii in two cases, optically thick \ce{HCO+} line emission and optically thick dust continuum emission:

\begin{enumerate}[\hspace{0.0cm}(a)]
\item If the \ce{HCO+} emission is optically thick, it prevents us to probe the inner dense region. An \ce{HCO+} hole would be seen in the \ce{HCO+} integrated intensity map due to line self-absorption and/or continuum subtraction. In order to reduce this effect, we integrated the spectra avoiding the optically thick regions near the systemic velocity (Table \ref{tab:map}) (see Appendix \ref{app:A}). 
Excluding the low-velocity channels should not affect the measured snowline location because it is expected to locate at the inner region where the velocity is high.
However, 
it is unclear what fraction of the continuum emission is absorbed by the foreground \ce{HCO+} gas at the selected velocity ranges for integration, resulting in an ``over subtraction'' in the continuum subtraction process.
This issue leads to a mis-identification of \ce{HCO+} depletion, e.g., as for the case of Per-emb-49 in Figure \ref{fig:hcop}. 
The current data cannot completely rule out this possibility except for sources with \ce{CH3OH} detections (see below).
\item If the dust continuum emission is optically thick at the frequency of interest, no line emission can escape from such a region. 
In this case, the \ce{HCO+} line emission will mimic the depletion signature due to the presence of the water snowline.
It is noteworthy that the dust opacity could be increased inside the \ce{H2O} snowline because the evaporation of icy grains leads to effective destructive collisions and higher dust densities \citep{ba15,ci16}.
If this is the case, the \ce{HCO+} depletion toward the center may still reflect the \ce{H2O} snowline locations
assuming that the optically thick dust region and an associated \ce{HCO+} depleted region are due to a dust opacity change within the water snowline.
\end{enumerate}

Since \ce{CH3OH} has a sublimation temperature similar to that of \ce{H2O} ($\sim100~{\rm K}$), 
\ce{CH3OH} line emission can be used as a proxy for the location of the \ce{H2O} snowline.
For those sources with \ce{CH3OH} detections, the measured \ce{HCO+} peak radii broadly agree with the \ce{CH3OH} emission extents (Figure \ref{fig:ch3oh}), which can resolve the issues of optical depths mentioned above.
This indicates that the estimates of the \ce{H2O} snowline locations from \ce{HCO+} are reasonable at least for these six sources.
Thus, we speculate that \ce{HCO+} is a good tracer of the \ce{H2O} snowline, but future observations with a resolution sufficient to resolve the continuum source or more warm-gas tracers are required to completely rule out the optical-depth issue.

\subsubsection{Dependence of the physical and chemical models}
\label{sec:model_depend}
The binding energy used in the chemical model determines the sublimation temperature, the decisive parameter for the snowline locations \citep{co03}. 
The binding energy of pure CO is found between $\sim850-1000$ K \citep{bi06} but can be increased to $\sim1200-1700$ K depending on the substrate \citep{fa16}, resulting in a CO sublimation temperature of $\sim17-33$ K at a gas density of $\sim10^7~{\rm cm^{-3}}$.
To make a model that fits the two burst sources (section \ref{sec:inburst}), Per-emb-27 and Per-emb-44, we use binding energies of 1307 K for CO (\citealt{no12}, measured from amorphous water ice) and 4820 K for \ce{H2O} \citep{sa93,fr01}.
Figure \ref{fig:lvsr} shows the modeled curve with the CO binding energy of 1150 K (\citealt{co04}, $T_{\rm sub}\sim21~{\rm K}$) and 1307 K ($T_{\rm sub}\sim25~{\rm K}$).
However, the binding energy is degenerate with the density structures (see below), and the latter are not necessarily the same in different sources.

Although a massive unstable disk is presumed to trigger the accretion burst, we perform our analysis using models without a disk.
A protostellar disk can shield the envelope or itself from the central radiation, changing the temperature structure mainly along the equatorial plane \citep{mu15}.
Figure \ref{fig:lvsr} (bottom) shows the modeled curve from a disk model in comparison with the no-disk model that is used for the \ce{HCO+} peak radii.
A higher central luminosity is required to heat the envelope and shift the \ce{H2O} snowline outward for the disk model compared with the no-disk model.
If a massive disk is included in the model, the number of post-burst sources and the burst luminosities are expected to increase.
However, a disk model is much more complicated since the disk density, geometry, and grain size distribution all affect the temperature structure.
To keep the model simple, we chose to use the no-disk model. 
The no-disk model could be considered as a conservative approach for identifying the occurrence of a past burst.
Specifically, the modeled snowline radii are an upper limit, and correspondingly the estimated burst luminosities, $L_{\rm burst,\ce{H2O}}$ and $L_{\rm burst,\ce{CO}}$, are lower limits.
MHW19 will discuss more details on how the disk size, geometry, envelope density and other parameters affect the temperature structures and snowline locations.


\section{Discussion}
\label{sec:discussion}

\subsection{Sources in the burst phase}
\label{sec:inburst}
Our sample includes two sources, Per-emb-27 (NGC 1333 IRAS2A) and Per-emb-44 (SVS 13A), which are likely undergoing an accretion burst given the current huge $L_{\rm bol}=30~L_\odot$ and $L_{\rm bol}=45~L_\odot$, respectively.

Several complex organic molecules are detected toward Per-emb-27 in the inner $40-100$ au region where the dust temperature is $>100$ K \citep{mar14,mau14,ta15}.
\citet{co14} found a knotty jet driven by Per-emb-27 with a dynamical time of $<30-90$ yr, which is considered to be a signature of episodic accretion \citep{vo18}. 
Per-emb-44 contains also a central hot region with a detection of glycolaldehyde \citep{de17,bi19}. Multiple components are found in continuum observations by \citet{to16,to18}, including a close binary with a separation of $0\farcs3$ (70 au).
Furthermore, \citet{le17} speculate the existence of a companion with a separation of $20-30$ au in order to explain the knotty jets with a period of $\sim300$ yr,
a putative companion that triggers the burst episodes in the close perihelion approach with an eccentric orbit.
These results suggest that Per-emb-27 and 44 are in the accretion-burst phase.

These burst-phase sources can be used as calibrators for the model.
If we assume that the current bolometric luminosity determines the observed CO and \ce{H2O} peak radii, the modeled curve in Figure \ref{fig:lvsr} should go through the data points of Per-emb-27 and 44.
As a result, the adopted model without a disk component and a CO sublimation temperature of $\sim25$ K looks reasonable (see Section \ref{sec:model_depend}).
However, the assumption is not necessarily true because the luminosity variation during a burst can be very large \citep{el16,vo18}.
Furthermore, the density structures should be quite different for each source.
Thus, this calibration provides only a rough confirmation.

\begin{figure}
\includegraphics[width=0.5\textwidth]{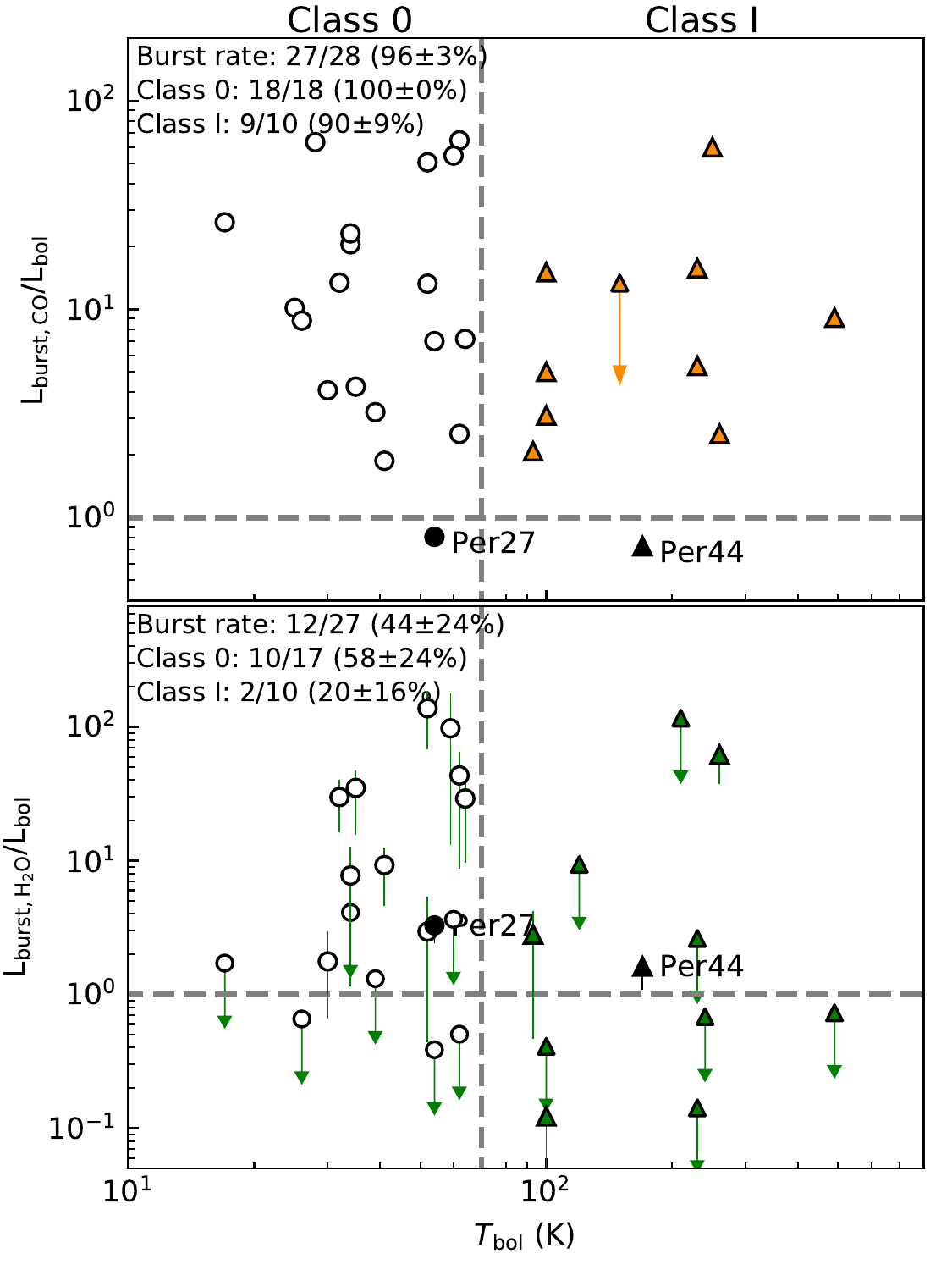}
\caption{$L_{\rm burst}/L_{\rm bol}$ as a function of the bolometric temperature $T_{\rm bol}$ from \ce{N2H+} (top, $L_{\rm burst, CO}/L_{\rm bol}$) and \ce{HCO+} (bottom, $L_{\rm burst, \ce{H2O}}/L_{\rm bol}$).
The markers shown in this plot is the same as that in Figure \ref{fig:lvsr} for Class 0 and Class I sources. The vertical dashed lines indicate the bolometric temperature of 70 K, the boundary between Class 0 and Class I. The horizontal dashed lines represent $L_{\rm burst}>L_{\rm bol}$ for classification of post-burst sources.}
\label{fig:tvsr}
\end{figure}

\subsection{Chronology of episodic accretion in protostars}
In this section we discuss the history of the episodic accretion process in a statistical way from Class 0 to Class I.
We derive frequencies of accretion bursts in Section \ref{sec:burst_fre} and mass accretion rates during burst phases in Section \ref{sec:macc}.
Then, based on these results, we discuss the mass accumulation history of protostars in episodic accretion in Section \ref{sec:acc_history}

\subsubsection{Evolution of burst frequency}
\label{sec:burst_fre}
Estimation of the outburst frequency has been done by monitoring a sample of protostars in the more evolved Class II stage \citep{sc13,co19}. These studies require a large survey with a long baseline in time \citep{hi15}. The chemical probes extend the baseline to 1000-10,000 yr by considering the refreeze-out time scales.

The refreeze-out time scales of \ce{CO} and \ce{H2O} are different because their snowlines are located at different radii with different densities.
Thus, these two chemical tracers provide complementary information to constrain the time since a past burst.
The refreeze-out time scales can be expressed as a function of gas density ($n_{\rm H_2}$) and dust temperature ($T_{\rm dust}$), i.e.,
\begin{equation}
\tau_{\rm fr}=1\times10^4~{\rm yr}\sqrt{\frac{10~\rm K}{T_{\rm dust}}}\frac{10^6~\rm cm^{-3}}{n_{\rm H_2}}
\vspace{4pt}
\end{equation}
from \citet{vi12} and \citet{vi15}.
Here we assume $\tau_{\rm fr}$ is 10,000 yr for CO and 1000 yr for \ce{H2O} \citep{vi15}.
As a result, we categorize these targets into: (1) post-burst source from \ce{HCO+}. The burst has occurred in the past 1000 yr. (2) post-burst source from \ce{N2H+} but not from \ce{HCO+}. The burst has occurred during the past 1000 to 10,000 yr. (3) no signature of a burst neither in \ce{N2H+} nor \ce{HCO+}. No burst has occurred during the past 10,000 yr (Table \ref{tab:sum}). 
Figure \ref{fig:prof4} shows the intensity profiles of the standard cases for these three categories together with that during an outburst.
Table \ref{tab:sum} lists the time since the last burst for each source.

\begin{figure}
\includegraphics[width=0.5\textwidth]{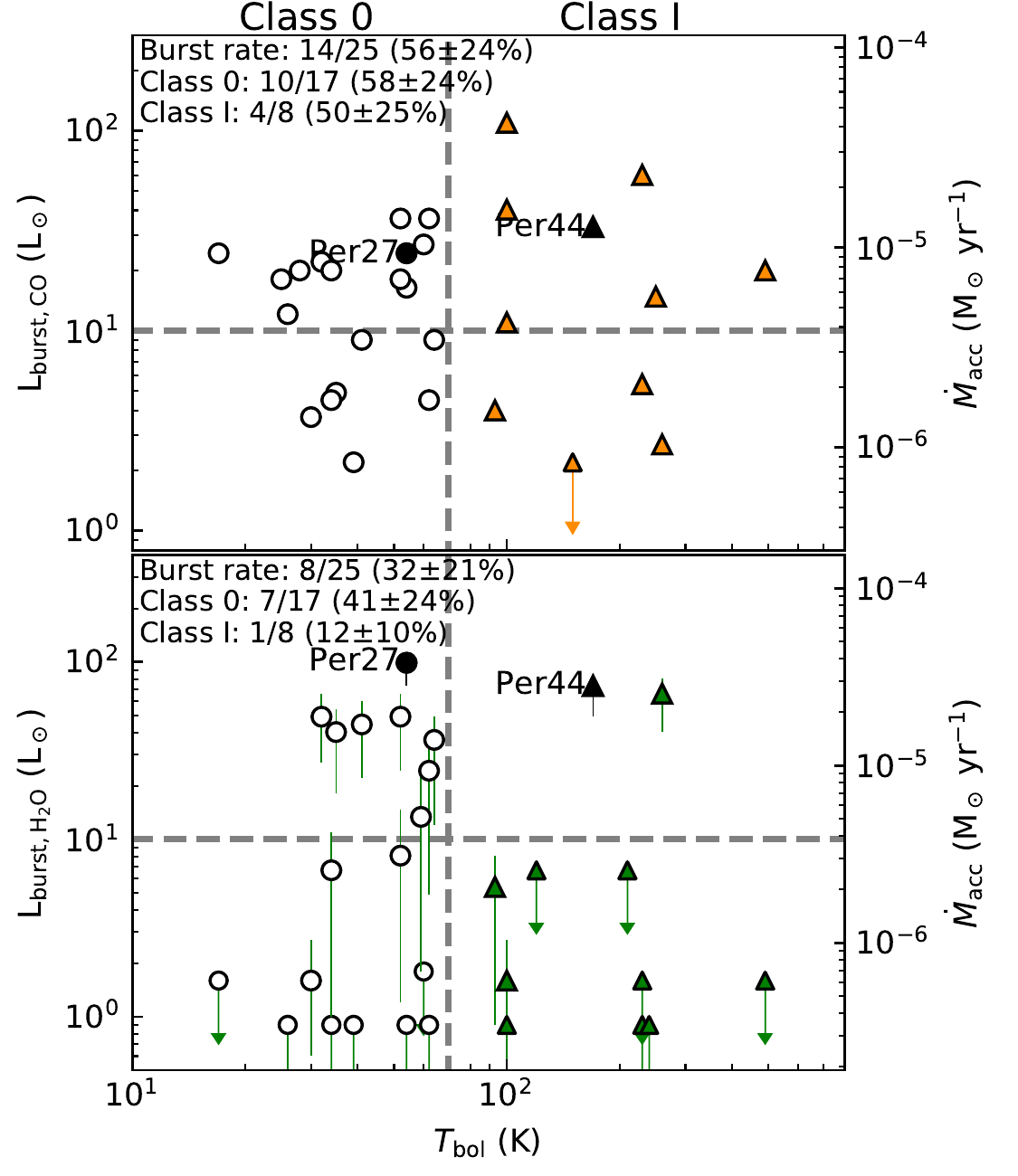}
\caption{Luminosity during the past burst obtained from \ce{N2H+} (top) and \ce{HCO+} (bottom) as a function of bolometric temperature. 
The luminosity is derived by modeling the line-emission peak offset, and sources with an offset less than the half-beam are labeled with the half-beam as an upper limit.
The horizontal dashed lines indicate the burst luminosity $L_{\rm burst}>10~L_\odot$.
The right axis shows the mass accretion rate derived assuming that the luminosity is dominated by accretion luminosity.}
\label{fig:tvsm}
\end{figure}

Figure \ref{fig:tvsr} shows $L_{\rm burst}/L_{\rm bol}$ as a function of the evolutionary indicator, bolometric temperature ($T_{\rm bol}$), from both CO and \ce{H2O}.
Sources with $L_{\rm burst}>L_{\rm bol}$ are identified as post-burst sources, namely that the source has experienced a past burst within the refreeze-out time.
Excluding the two sources in the burst phase (Per-emb-27 and 44, section \ref{sec:inburst}), there are 28 and 27 sources for the following statistical analyses with \ce{N2H+} and \ce{HCO+}, respectively.
$L_{\rm burst}$ is defined as an upper limit for those sources whose measured peak radius is less than the half-beam size, and the upper limit is obtained using the half-beam size as the peak radius.
As a result, we cannot identify the chemical signature of a past burst for Per-emb-51 on the basis of its \ce{N2H+} map, and for seven sources on the basis of their \ce{HCO+}
maps.
For Per-emb-51 with \ce{N2H+}, the upper limit of $L_{\rm burst, \ce{CO}}$, $2.2~L_\odot$, suggests that it unlikely experienced a past burst, or at least a strong burst.
For those seven sources from $L_{\rm burst, \ce{H2O}}$, depending on the weighting of the map, the upper limits of $L_{\rm burst, \ce{H2O}}$ are $0.9~L_\odot$ for three sources, $1.8~L_\odot$ for two sources, and $6.7~L_\odot$ for two sources. 
These upper limits are generally smaller than the burst luminosities of the post-burst sources, a median of $31.1~L_\odot$ and a standard deviation of $20.8~L_\odot$ (Figure \ref{fig:tvsm}).
Therefore, we do not consider these sources to be post-burst sources.
As a result, we find $100\pm0$\% of Class 0 objects are post-burst sources and $90\pm9$\% of Class I objects are post-burst sources from \ce{N2H+} alone, where the uncertainty is derived using binomial statistics.
This result implies that the burst interval is $<$10,000 yr for Class 0 objects and $\sim$11,000 yr ($\frac{10,000}{0.9}$ yr) for Class I objects.
From \ce{HCO+} alone, we identify $58\pm24$\% of Class 0 objects and $20\pm16$\% of Class I objects are post-burst sources.
This gives us burst intervals of $\sim$1700 yr for Class 0 objects and $\sim$5,000 yr for Class I objects.
The combination of these results suggests that the burst frequency is decreasing from the Class 0 to the Class I stage.

The criterion used to define post-burst sources, $L_{\rm burst}>L_{\rm bol}$, may not well portray sources during an accretion burst especially for the estimate of burst frequencies; intuitively, large bursts occur rarely compared with small bursts. 
For example, several post-burst sources (Per-emb-3, 9, 30, etc.) have their $L_{\rm burst, CO}$ of $\sim2.2-4.5~{L_\odot}$, only $\sim3-4$ times larger than their $L_{\rm bol}$.
These sources may have experienced a small accretion outburst ($\dot{M}_{\rm acc}\sim(1-2)\times10^{-6}~M_\odot~{\rm yr^{-1}}$, see section \ref{sec:macc}) that is expected to occur more frequently.
In addition, the observed bolometric luminosity can be affected by the viewing angle of a disk-outflow system. A small $L_{\rm bol}$ may result from a nearly edge-on configuration \citep{of12}. 
This can lead to a misidentification of a post-burst source if $L_{\rm bol}$ is very small and if $L_{\rm bust}$ is only slightly larger.
Thus, we decide to focus on those post-burst sources robustly identified.
If we consider only large outbursts with $L_{\rm burst}>L_{\rm bol}$ and $L_{\rm burst}>10~L_\odot$ (\citealt{en09}),
$56\pm24$\% of Class 0 objects and $50\pm25$\% of Class I objects are post-burst sources from \ce{N2H+}, implying an interval of $\sim$18,000 yr and $\sim$20,000 yr for Class 0 and Class I, respectively (Figure \ref{fig:tvsm}).
From \ce{HCO+}, there are $41\pm24\%$ of post-burst sources in Class 0 with an interval of $\sim$2,400 yr and $12\pm10\%$ of post-burst sources in Class I with an interval of $\sim$8,000 yr for Class I (note that two sources with $L_{\rm bol}>10~L_\odot$ are excluded given the new criterion).
The inconsistencies between the burst intervals traced with \ce{N2H+} and \ce{HCO+} is difficult to explain but this result still suggests a decrease of burst frequency from the Class 0 stage to the Class I stage.
However, from an evolutionary point of view, if a disk is significantly denser/larger at the Class I stage than that at the Class 0 stage, it might shrink the emission peak inward (Figure \ref{fig:lvsr}). 
The disk evolution may thus lead us to underestimate the number of post-burst sources at the Class I stage, and our conclusion that the burst frequency decreases from the Class 0 to the Class I stage might in turn not be robust.

Based on \ce{N2H+} observations, \citet{hs18} found that episodic accretion can start at a very early evolutionary stage.
Here we find that the burst frequency is higher in the Class 0 stage than in the Class I stage.
The accretion outbursts are believed to be associated with a massive and large disk with the gravitational and/or magnetorotational instability \citep{vo15,zh10b}. 
Therefore, the burst frequency may reflect the disk formation and/or evolution.
The onset of disk formation is still unclear, but disks have been found in some Class 0 sources \citep{to12,mu13,oh14,le17b,le18,as17,hs19,ma19}.
Besides, numerical simulations suggest that an initially unstable cloud core can promote disk formation and tend to have a higher mass accretion rate \citep{ma16}.
\citet{vo13} further support that the episodic accretion process is highly dependent on the core initial conditions;
with a higher ratio of rotational to gravitational energy, the strength of the burst is increased.

If accretion bursts are triggered by infalling fragments in an gravitationally unstable disk \citep{vo05}, the decreasing burst-frequency implies that, at an earlier stage, either disk fragmentation occurs more frequently or that the fragments tend to fall more often onto the central source. 
It is also noteworthy that \citet{re17} find that the gravitational instability in \citet{vo05} is overestimated with a fixed central source because the disk angular momentum can translate into the orbital motion of the central source.
The gravitational instability can be controlled by infall onto the disk and its thermodynamics \citep{kr16}. 
A high mass infall rate is crucial for sustaining the gravitational instability in disks \citep{vo05,kr08}, which might explain the high burst frequency in the Class 0 stage.
In addition, fragmentation is suggested to occur in cold regions which require a sufficient cooling time \citep{vo10,kr10,kr11}. 
Observations of multiple systems \citep{mu16,to18} in the cold disk/envelope support such fragmentations at an early stage.

\subsubsection{Mass accretion rate}
\label{sec:macc}
The derived burst luminosities give us indications on the mass accretion rate during the past burst phase.
If we assume that $L_{\rm burst}=L_{\rm acc}$, the accretion luminosity, we derive the mass accretion rate with
\begin{equation}
L_{\rm acc}=\frac{G M_{\rm star} \dot{M}_{\rm acc}}{R}
\end{equation}
where $G$ is the gravitational constant, $R$ is the protostar radius (assumed to be 3 $R_\odot$, \citealt{du10}), and $M_{\rm star}$ is the mass of the central source (assumed to be $0.25~M_\odot$, \citealt{ev09}, half of the average stellar mass).
Figure \ref{fig:tvsm} shows the inferred burst-phase mass accretion rate as a function of the evolutionary indicator $T_{\rm bol}$.
This figure aims to reveal the evolution of episodic accretion while the ratios of the post-burst sources ($>10~L_\odot$) indicate the burst frequency, and $\dot{M}_{\rm acc}$ represents their burst strength.
We find mass accretion rates during the burst phase ($L_{\rm burst}>L_{\rm bol}$ and $L_{\rm burst}>10~L_\odot$) between $4.2\times10^{-6}~{M_\odot~{\rm yr}^{-1}}$ and $4.2\times10^{-5}~{M_\odot~{\rm yr}^{-1}}$ from \ce{N2H+} and between $5.1\times10^{-6}~{M_\odot~{\rm yr}^{-1}}$ and $2.5\times10^{-5}~{M_\odot~{\rm yr}^{-1}}$ from \ce{HCO+} (Table \ref{tab:sum}).
The median is $\sim7.6\times10^{-6}~{M_\odot~{\rm yr}^{-1}}$ from \ce{N2H+} and $\sim1.6\times10^{-5}~{M_\odot~{\rm yr}^{-1}}$ from \ce{HCO+} with a standard deviation of $\sim(5.8-8.8)\times10^{-6}~{M_\odot~{\rm yr}^{-1}}$.
This systematic shift might come from the adopted model parameters such as the CO binding energy.
From an evolutionary point of view, the median does not change from Class 0 to Class I from both \ce{N2H+} and \ce{HCO+}.
However, the estimation has a strong bias in selection as the analyses only identify those sources with strong bursts.
Besides, the assumption of $M_{\rm star}=0.25~M_\odot$, can be unrealistic as the stellar mass must increase from the Class 0 to the Class I stage.
Thus, $M_{\rm star}$ should be an increasing value as a function of time, and given the similar accretion luminosity during outbursts from Class 0 to Class I, $\dot{M}_{\rm acc}$ is subsequently expected to decrease with time.

\subsubsection{Mass accumulation of protostars}
\label{sec:acc_history}
We have derived the mass accretion rates and the intervals between accretion episodes, which allows us to probe the growing process of the central stars.
Considering a lifetime of $0.15-0.24$ Myr \citep{du15} and an interval of 2400 yr for the Class 0 stage, accretion bursts would occur $63-100$ times during this stage. 
Similarly in Class I, with a lifetime of $0.31-0.48$ Myr and an interval of 8,000 yr, accretion bursts would occur $39-60$ times; 
note that our sample includes only one Late Class I protostar (Per-emb-63) with $T_{\rm bol}>300$ K \citep{ev09}, and the burst frequency likely keeps decreasing in the Class I stage; thus, the total burst number might be overestimated in Class I.
Besides, if the lifetimes of Class 0 and Class I are 30\% shorter as suggested by \citet{car16}, the total number of bursts would be revised downward by 30\%.

Statistically, \citet{en09} found that $\sim5\%$ of the embedded protostars have a high luminosity with $L_{\rm bol}>10~L_\odot$, or $\dot{M}_{\rm acc}>10^{-5}~M_\odot~{\rm  yr}^{-1}$.
If these sources are during the burst phase, each burst would last for $2400\times0.05=120$ yr at Class 0 stage, which is consistent with the duration of $100-200$ yr predicted from simulations of \citet{vo05}.
Given the median burst-phase accretion rate ($7.6-16.2)\times10^{-6}~{M_\odot~{\rm yr}^{-1}}$, each burst would thus deliver $\sim9.1-19.4\times10^{-4}~{M_\odot}$ onto the central star.
Thus, the protostar would accumulate $\sim0.06-0.19~{M_\odot}$ at the Class 0 stage and $\sim0.04-0.12~{M_\odot}$ at the Class I stage.
Assuming a mean stellar mass of $0.5~M_\odot$ \citep{ev09}, this result implies that only $\sim20-60\%$ ($0.1-0.3~M_\odot$) of mass is accumulated during burst phases.
A simple explanation for the discrepancy is that the final stellar mass is in fact smaller;
for example, the peak of the initial mass function is $\sim0.3~M_\odot$ \citep{mu02,al07}.
Alternatively, we propose three possibilities to complement the remaining mass accumulated:
(1) an underestimation of the burst-phase mass accretion rate. The accretion rate is estimated considering a model without a disk component such that the derived accretion luminosity is likely a lower limit. 
(2) non-negligible mass accumulation during the quiescent phase. 
As our estimated accumulated mass of $0.1-0.3~M_\odot$ in Class 0/I stage arises solely from the accretion burst, it requires $\lesssim80\%$ ($0.4~M_\odot$) of the mass accreted in quiescent phase with $\dot{M}_{\rm acc}\lesssim(5.6-8.7)\times10^{-7}~M_\odot~{\rm yr}^{-1}$ to build a star with an average mass of $0.5~M_\odot$.
(3) the existence of super bursts (i.e. FU ori-type). 
\citet{of11} estimate that $\sim25\%$ of mass is accreted during the FU ori events with $\dot{M}_{\rm acc}\sim10^{-5}-10^{-4}~{\rm M_\odot~yr^{-1}}$.
This is consistent with the highest mass accretion rate estimated from the \ce{N2H+} observations as $\sim4.2\times10^{-5}~{\rm M_\odot~yr^{-1}}$. If such super bursts last for longer, they could deliver significant material onto the central sources.

\section{Summary}
\label{sec:summary}
We present our ALMA cycle 5 observations of \ce{N2H+} ($1-0$) and \ce{HCO+} ($3-2$) toward 39 Class 0 and Class I sources.
We analyze the spatial distributions of these two molecules, and by comparing to our chemical models, we derive the required luminosity that sublimates \ce{CO} and \ce{H2O} and destroys \ce{N2H+} and \ce{HCO+}, respectively.
We compare such derived luminosity to the bolometric luminosity (the current luminosity), thus identifying the sources that experienced a past accretion burst with $L_{\rm burst}>L_{\rm bol}$, i.e., the post-burst sources.
Our results are summarized as follows: 
\begin{enumerate}
\item \ce{N2H+} and \ce{HCO+} peak positions can be used to trace the \ce{CO} and \ce{H2O} sublimation regions, respectively, and in turn to estimate the luminosity during the past burst. 
While \ce{N2H+} at large scale is less affected by the system geometry, \ce{HCO+} 
in the inner regions is sensitive to the inclination angle but is crucial to trace the past burst over a shorter timescale.
\item We find that 7/17 Class 0 and 1/8 Class I are post-burst sources from \ce{HCO+}. 
This decrease of the fraction of post-burst sources may result from the evolution of burst frequency, but we cannot exclude the possibility that the snowline radius is shrunk due to the increase of disk density/size from the Class 0 to the Class I stage.
If the disk evolution is not the main factor, then we can draw the following conclusions about the mass accumulation history from the Class 0 to the Class I stage.
\item We derive the intervals between accretion episodes of $\sim2,400$ yr for Class 0 sources and $\sim8,000$ yr for Class I sources, suggestive of a decrease in burst frequency during the embedded phase. 
If the accretion outburst is triggered by disk fragmentation due to gravitational instability, our result suggests that the fragmentation occurs more frequently at an earlier evolutionary stage.
Alternatively, the fragment has a higher probability to fall onto the central star at such a stage.
\item We estimate the mass accretion rates at the burst-phase to be $(7.6-16.2)\times10^{-6}~{\rm M_\odot~yr^{-1}}$. From an evolutionary point of view, the burst magnitude is likely unchanged from Class 0 to Class I.
\item Based on the estimate of mass accretion rate and interval between episodes, we derive an accumulated mass of $0.06-0.19~{\rm M_\odot}$ at the Class 0 stage and $\sim0.04-0.12~{M_\odot}$ at the Class I stage, in total $0.1-0.3~{M_\odot}$ during burst phases.
This value is smaller than the typical stellar mass of $0.3-0.5~{M_\odot}$.
More material needs to be accreted to build the star during the quiescent phase or perhaps the star can accumulate mass via a few super accretion bursts.
\end{enumerate}



We are thankful for the referee for many insightful comments for the discussion that helped to improve this paper significantly.
The authors thank Merel van 't Hoff, Jeong-Eun Lee, and Marc Audard for providing valuable discussions.
This paper makes use of the following ALMA data:\ ADS/JAO.ALMA\#2017.1.01693.S. ALMA is a partnership of ESO (representing its member states), NSF (USA) and NINS (Japan), together with NRC (Canada), MOST and ASIAA (Taiwan), and KASI (Republic of Korea), in cooperation with the Republic of Chile. The Joint ALMA Observatory is operated by ESO, AUI/NRAO and NAOJ.
T.H.H. and N.~H. acknowledges the support by Ministry of Science and Technology of Taiwan (MoST) 107-2119-M-001-041 and 108-2112-M-001-048.
N.H. acknowledges a grant from MoST 108-2112-M-001-017.
C.W. acknowledges financial support from the University of Leeds and from the Science Facilities and Technology Council (STFC), under grant number ST/R000549/1.
J.K.J. acknowledges support from the European Research Council (ERC) under the European Union's Horizon 2020 research and innovation programme (grant agreement No~646908).
S.P.L. acknowledges the support from the Ministry of Science and Technology (MOST) of Taiwan with grant MOST 106-2119-M-007-021-MY3

{{\it Software:} CASA \citep{mc07}, RADMC-3D \citep{du12}, Astropy \citep{as13}, APLpy \citep{ro12}}

\appendix

\setcounter{figure}{0}
\setcounter{table}{0}
\renewcommand{\thefigure}{A\arabic{figure}}

\section{A. Comparison of continuum subtracted \ce{HCO+} maps}
\label{app:A}
\begin{figure*}
\includegraphics[width=\textwidth]{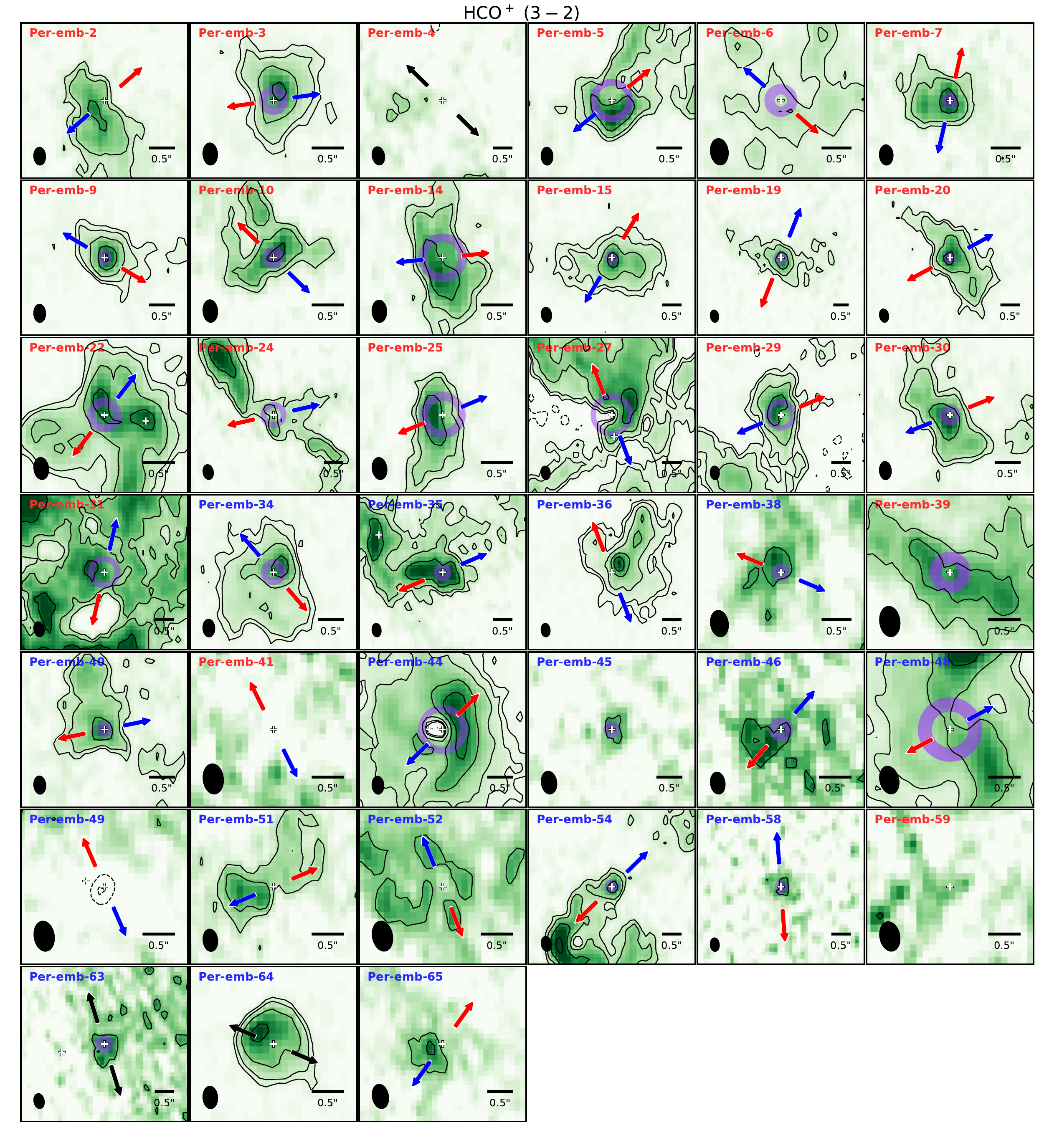}
\caption{Same as Figure \ref{fig:n2hp} but for HCO$^{+}$ integrated intensity maps. The contour levels are 3$\sigma$, 5$\sigma$, 10$\sigma$, 20$\sigma$, 30$\sigma$, 40$\sigma$ with $\sigma$ the rms noise level given in Table 2. 
The black bar in the lower right corner indicates a size of 0\farcs5 while the size of the image is adjusted panel by panel.
The purple circles represent the radii of the measured \ce{HCO+} peak with the thickness for representing the uncertainties (see the text for detail).}
\label{fig:hcop}
\end{figure*}

\begin{figure*}
\includegraphics[width=0.9\textwidth]{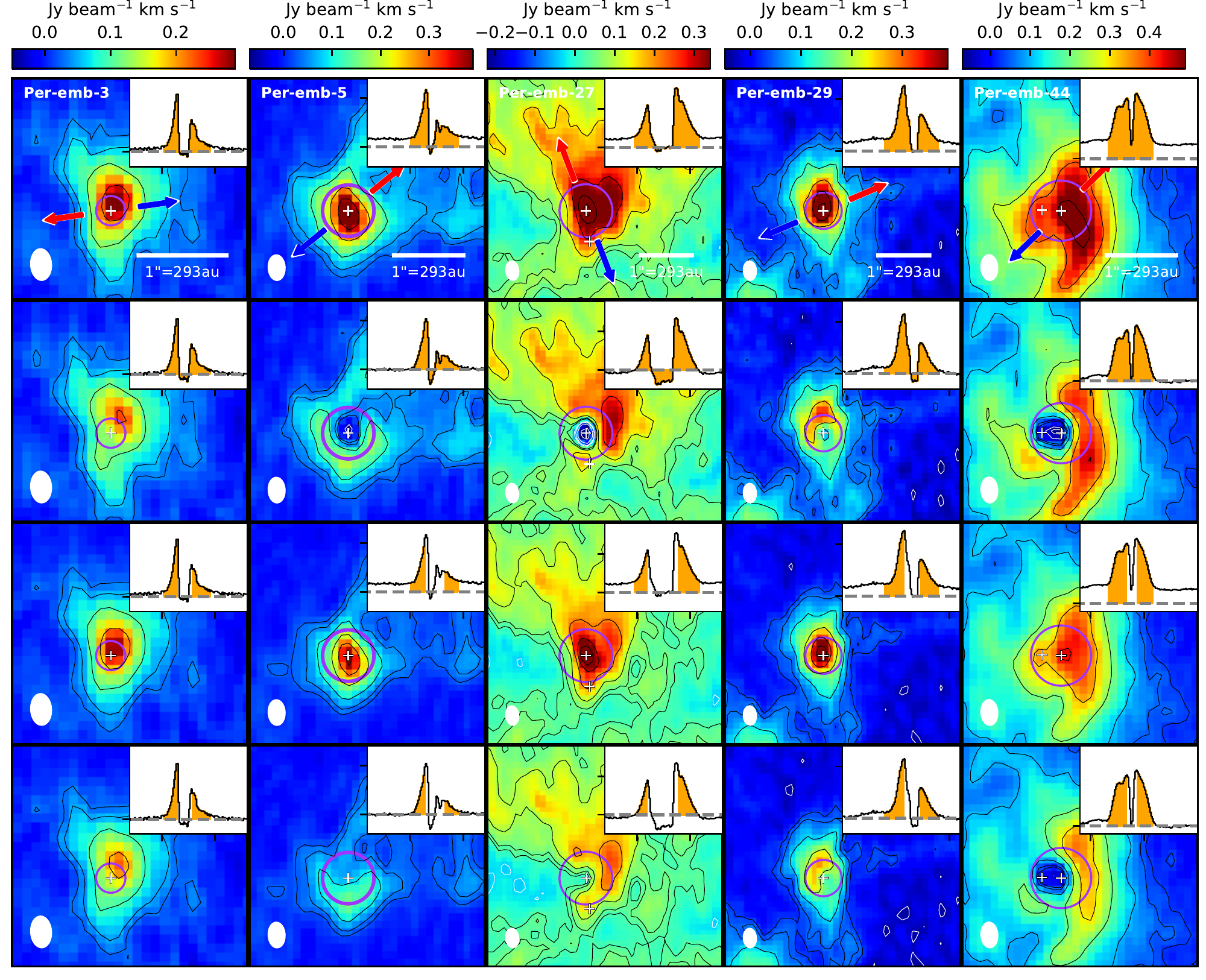}
\caption{Comparison of \ce{HCO+} integrated intensity maps with different velocity ranges with and without continuum subtraction. 
The panel from top to the bottom are (1) normal integration without continuum subtraction, (2) normal integration with continuum subtraction, (3) integration in the optically thin region without continuum subtraction, and (4) integration in the optically thin region with continuum subtraction.
The contour levels are $\pm3\sigma$, $\pm5\sigma$, $\pm10\sigma$, $\pm20\sigma$, $\pm30\sigma$, $\pm40\sigma$ as the positive (negative) values are shown in black (white). The insert in each panel shows the spectrum and the integration ranges in orange. The fluxes are on arbitrary scales, and the grey dashed line marks the zero level. The purple circles indicate the radii of the measured peak positions.}
\label{fig:hcop_com}
\end{figure*}

Figure \ref{fig:hcop} shows the \ce{HCO+} integrated intensity maps toward the 39 targets. The selected velocity ranges for integration are listed in Table \ref{tab:map}.
To study the influences from these selection, Figure \ref{fig:hcop_com} shows the images of five selected targets with four types of maps for comparison which are (1) normal integration without continuum subtraction, (2) normal integration with continuum subtraction, (3) integration in the optically thin region without continuum subtraction, and (4) integration in the optically thin region with continuum subtraction.
The emission peaks in the type (1) maps are mostly toward the source center.
However, with continuum subtraction shown in type (2), a hole appears at the center and three of them have negative values. 
Such negative values come from the ``over subtraction'' for the continuum emission (or strong line absorption) when a significant fraction of the continuum emission is absorbed by the foreground molecular gas.
Thus, this hole does not properly reflect the \ce{HCO+} spatial distribution.
We therefore integrated the flux avoiding the optically thick region near the central velocity which is shown in type (3); we integrated the velocity ranges excluding the central channels in between the peak of the blue and red-shifted emission (Figure \ref{fig:hcop_com} and Table \ref{tab:map}).
As a result, after continuum subtraction, type (4), the negative contours disappear toward the central region.
This process still cannot completely remove the over subtraction because the continuum emission is also absorbed by molecular gas at the selected velocity range.
However, the fraction of the absorbed continuum emission should not be significant if the selected velocity range is optically thin. 
Thus, this process likely minimizes the issue of over subtraction.

\setcounter{figure}{0}
\renewcommand{\thefigure}{B\arabic{figure}}
\section{B. Modeled images and their measured peak radii}
\label{app:B}
\begin{figure*}
\centering 
\includegraphics[width=0.88\textwidth]{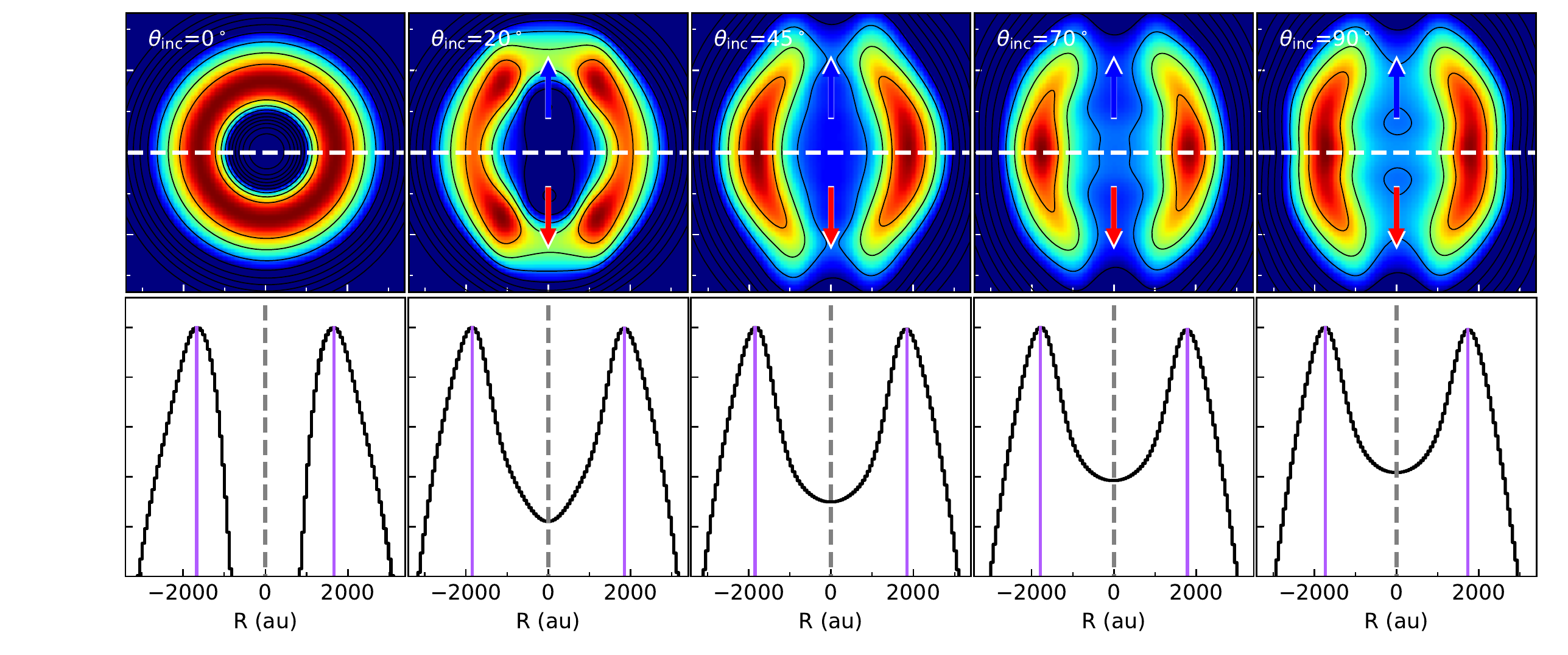}
\vspace{-10pt}
\caption{({\it top}) \ce{N2H+} ($1-0$) modeled images at different inclination angles ($0\arcdeg$ for pole-on) for a source with $L=30~L_\odot$. The images are normalized to the peak value with a contour step of 10\%. The horizontal dashed lines shows the cut used for the intensity profile in the bottom panel.
({\it bottom}) Intensity profiles along a horizontal cut for each modeled image. The x-axis is in the same scale as the above image for comparison. The dashed line represents the source position. The purple vertical lines indicate the measured peak positions that are used to compare with the observed images.}
\label{fig:n2hp_peak}
\end{figure*}

\begin{figure*}
\centering 
\includegraphics[width=0.88\textwidth]{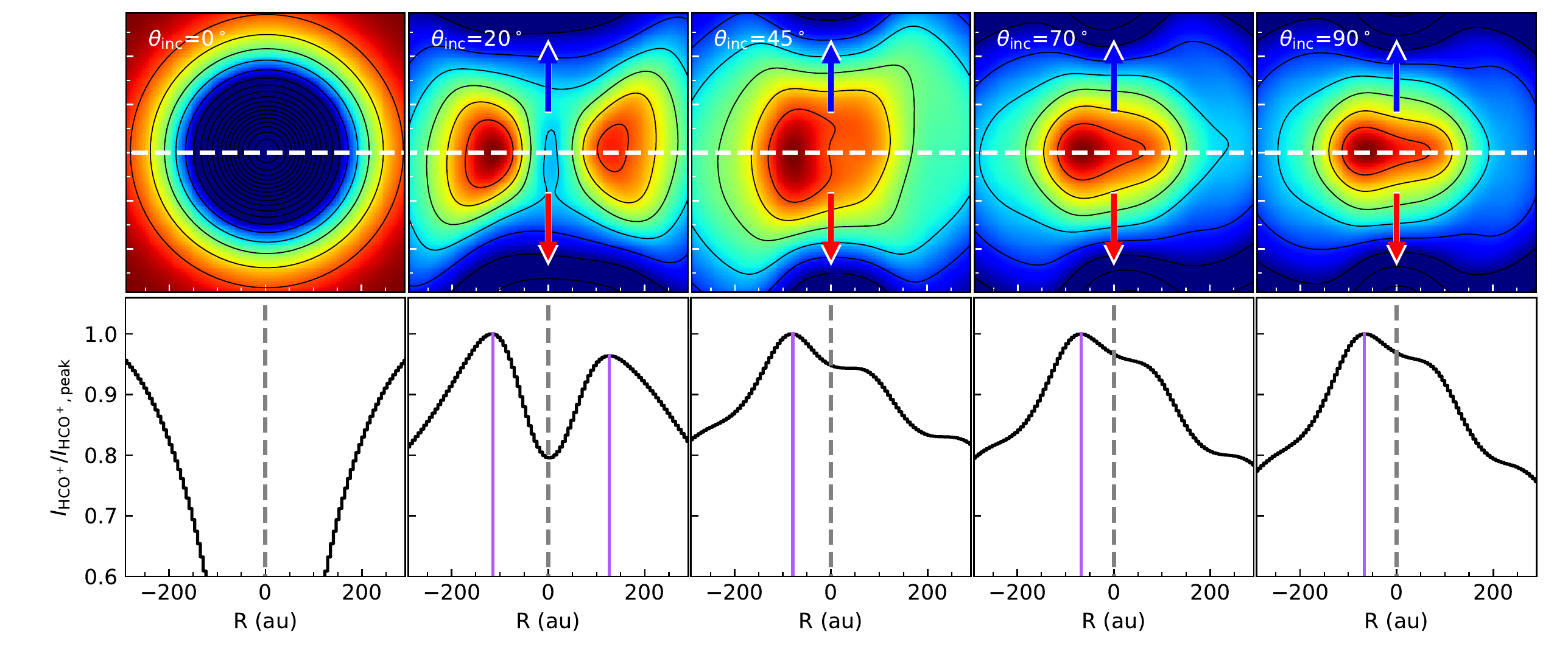}
\vspace{-10pt}
\caption{Same as Figure \ref{fig:n2hp_peak} but for \ce{HCO+} ($3-2$) with a contour step of 5\%.}
\label{fig:hcop_peak}
\end{figure*}

\begin{figure*}
\centering 
\includegraphics[width=0.78\textwidth]{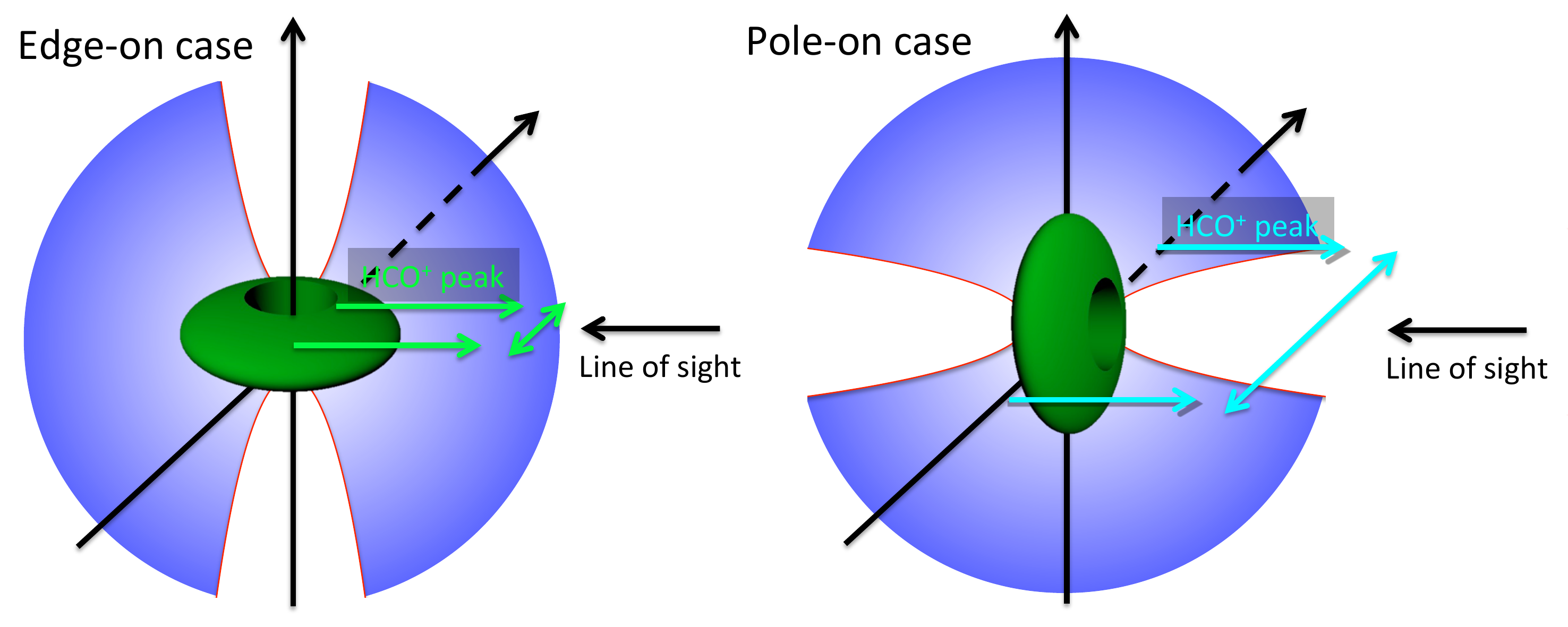}
\vspace{-10pt}
\caption{ Schematic illustration for the shift of \ce{HCO+} peak positions from the edge-on to pole-on case. This scenario suggests that the \ce{HCO+} emission is dominated by the inner region in edge-on case and by the outer envelope in the pole-on case.}
\label{fig:sch}
\end{figure*}

To compare with observations, we make \ce{N2H+} and \ce{HCO+} images from our chemical models.
We assume a Keplerian motion within 100 au around a protostar with a central mass $M_{\rm star}=0.25~M_\odot$.
At a radius beyond 100 au, the envelope follows the conservation of angular momentum and freefall with a rotation velocity $\propto r^{-1}$ and a radial velocity $\propto r^{-0.5}$.
Together with the given temperature, density and molecular abundance, we use the line radiative transfer code, available in RADMC3D \citep{du12}, to produce the images.
We make the images with inclination angles of 15\arcdeg, 25\arcdeg, 45\arcdeg, 65\arcdeg, and 75\arcdeg and convolve them by a gaussian with an FWHM of $1\farcs93$ for the \ce{N2H+} ($1-0$) maps and $0\farcs27$ for the \ce{HCO+} ($3-2$) maps, about the equivalent width of the observational beams.
To demonstrate inclination effects, here we show an example with $L=30~L_\odot$ at five inclination angles including the edge-on and pole-on cases  (Figures \ref{fig:n2hp_peak} and \ref{fig:hcop_peak}).
Although the inclination angle does not significantly change the peak radius of \ce{N2H+}, it influences that of \ce{HCO+} on small scales.
At the nearly pole-on configuration ($\theta_{\rm inc}\lesssim20\arcdeg$), the peak of the \ce{HCO+} emission is associated with gas at a higher latitude rather than that in the midplane; thus, it is dominated by the large-scale envelope rather than the inner \ce{H2O} snow line location (Figure \ref{fig:sch}).
In such a case, the \ce{HCO+} peak positions are located at a larger radius depending on the outflow opening angle.
In addition, toward larger inclination angles, the redshifted component from the rotating inner envelope can be absorbed by foreground infalling core such that only the blue-shifted component is left.
This opacity influence is expected to depend on the input density and the input velocity field.
We note that the modeled images have negligible dust continuum emission.
Thus, the ``over subtraction'' issue does not exist, and we integrate the full line profiles (appendix \ref{app:A}).
It is also noteworthy that the peak position of the integrated intensity map can change with a different selected velocity ranges;
the velocity range will decide the emitting regions that contribute to the map depending on the inclination angle.
As a result, the modeled curves of emission peak as a function of luminosity are shown in Figure \ref{fig:lvsr} in comparison with observations.
Because of the convolution, a transition close to the half-beam size is shown in Figure \ref{fig:lvsr}, denoting the limit of the resolution in our observation for resolving the central depletion.

\setcounter{figure}{0}
\renewcommand{\thefigure}{C\arabic{figure}}
\section{C. Intensity profiles}
\label{app:C}
\begin{figure*}
\centering 
\includegraphics[width=0.9\textwidth]{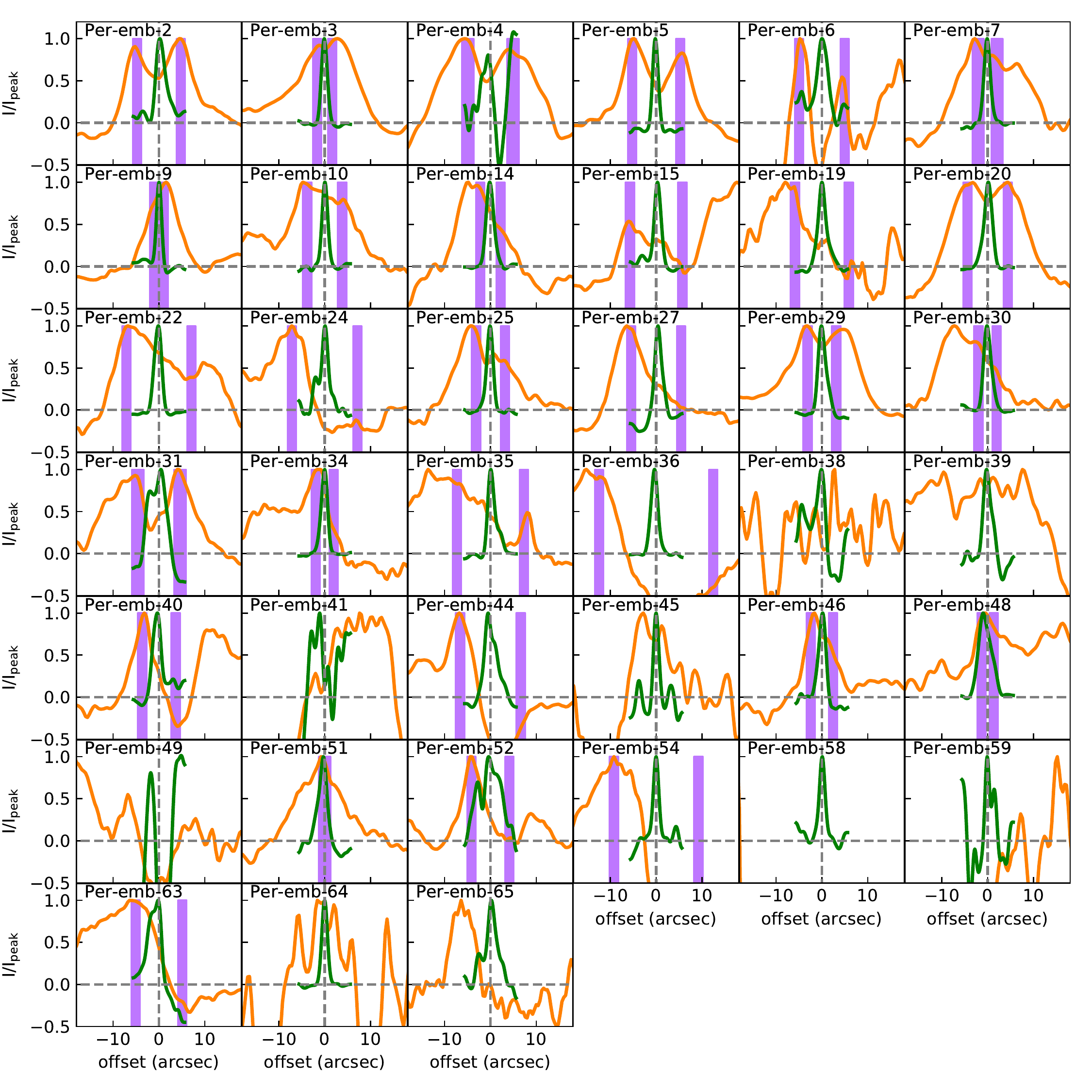}
\caption{ Same as the top panel of Figure \ref{fig:prof4} (\ce{N2H+} and \ce{HCO+}) but for all of the sources.}
\label{fig:n2hp_prof}
\end{figure*}

\begin{figure*}
\centering 
\includegraphics[width=0.9\textwidth]{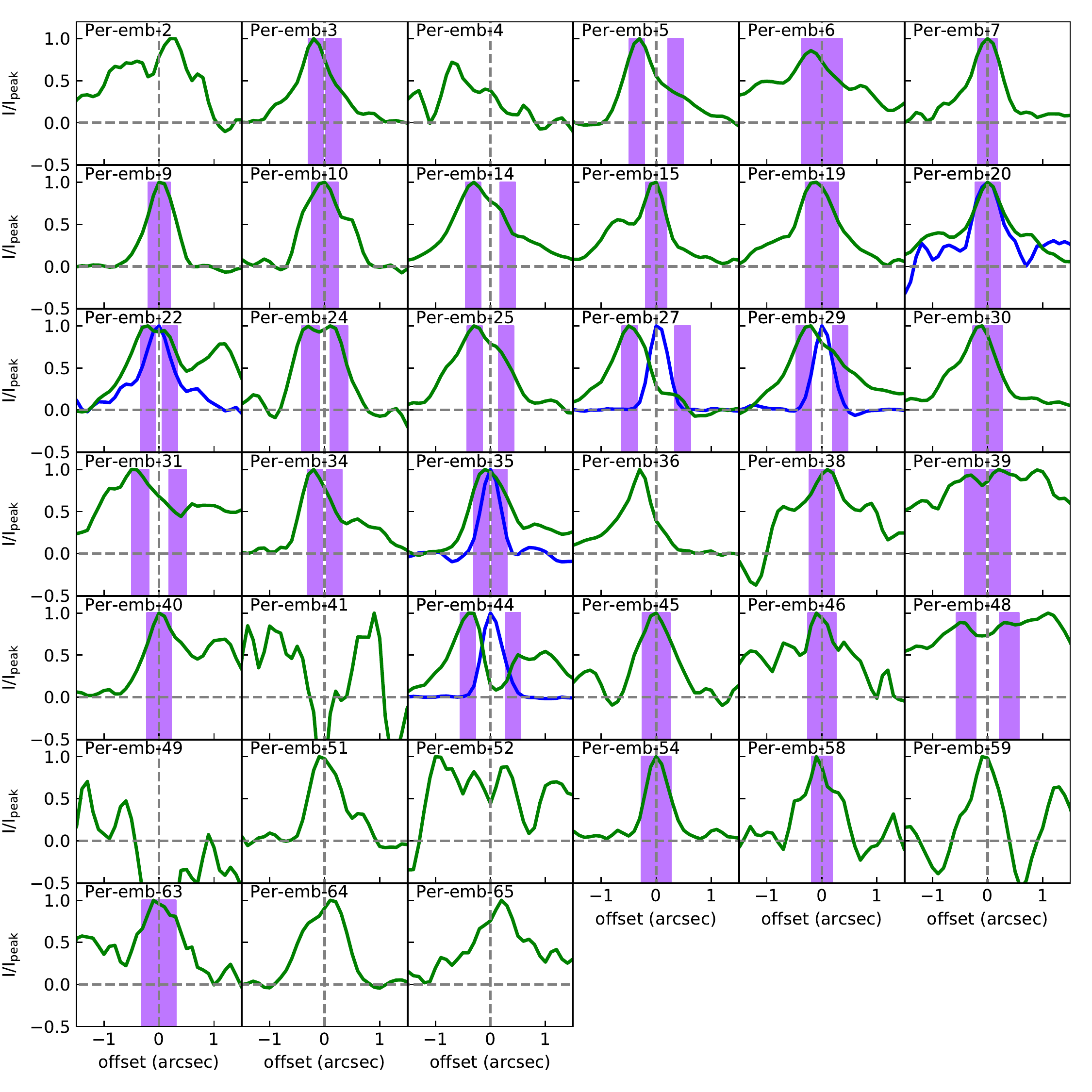}
\caption{ Same as the bottom panel of Figure \ref{fig:prof4} (\ce{HCO+} and \ce{CH3OH}) but for all of the sources.}
\label{fig:hcop_prof}
\end{figure*}

To check the correlation between the two pairs of molecules, \ce{N2H+}-\ce{HCO+} and \ce{HCO+}-\ce{CH3OH}, we plot the intensity profiles across the source centers and the measured peaks from Section \ref{sec:peak} (Figures \ref{fig:n2hp_prof} and \ref{fig:hcop_prof});
\ce{CH3OH} is plotted in the only six sources with detections toward the center. 
Anti-correlation between these pairs of molecules are found in most of the cases, and the measured peak positions are reasonable.

\end{document}